\documentclass[a4paper,12pt,onecolumn]{article}
 \usepackage[utf8]{inputenc} 
\usepackage[english]{babel}

\usepackage{fancyhdr} 
\usepackage[pdftex]{graphicx} 
\usepackage{url} 
\usepackage{caption} 
\usepackage{subcaption} 
 \usepackage{natbib} 
\usepackage{multirow} 
\usepackage{float} 
\usepackage{amssymb}
\usepackage[intlimits]{amsmath} 
\usepackage{amsthm} 
\usepackage[german]{nomencl}
\usepackage{booktabs} 
\usepackage{geometry}
\usepackage{enumitem}
\usepackage{array}
\newcolumntype{C}[1]{>{\centering\let\newline\\\arraybackslash\hspace{0pt}}m{#1}}
\makenomenclature
\usepackage[pdfborder={000},colorlinks=true,linkcolor=blue,citecolor=blue]{hyperref}
\usepackage{listings}
\usepackage{soul}

\usepackage[usenames,dvipsnames]{color}
\usepackage{tikz}
\usetikzlibrary{shapes,arrows,shadows, positioning}
\usepackage{bm,times}
\usepackage{smartdiagram}

\usepackage{attachfile}

\usepackage{lipsum}

\usepackage{arydshln}

\usepackage{footnote}
\makesavenoteenv{tabular}

\newcommand\blfootnote[1]{%
  \begingroup
  \renewcommand\thefootnote{}\footnote{#1}%
  \addtocounter{footnote}{-1}%
  \endgroup
}

\definecolor{b}{rgb}{0,0,.8}	
\definecolor{g}{rgb}{0,.6,0}	
\definecolor{n}{rgb}{0,0,0}	
\definecolor{h}{rgb}{0.4,0.2,0.2}	
\definecolor{v}{rgb}{0.2,0.6,0}

\newcommand{\I}{{\mathbb I}}

\newcommand{\DD}{{\mathcal{D}}}

\newcommand{\II}{{\mathcal{I}}}
\newcommand{\JJ}{{\mathcal{J}}}
\newcommand{\KK}{{\mathcal{K}}}

\newcommand{\bsU}{\boldsymbol U}

\newcommand{\bsX}{\boldsymbol X}
\newcommand{\bsY}{\boldsymbol Y}
\newcommand{\bsZ}{\boldsymbol Z}

\newcommand{\bsbeta}{\boldsymbol \beta}

\newcommand{\eps}{{\varepsilon}}

\DeclareMathOperator*{\argmin}{arg\,min}

\newcommand{\ov}\overline
\newcommand{\what}{\widehat}

\newcommand{\rig}\right
\newcommand{\lef}\left
\newcommand{\nf}\normalfont

\definecolor{rickgreen}{rgb}{0,0.6,0}

\definecolor{strikeout}{rgb}{0.5,0.5,.5}
\definecolor{newchange}{rgb}{0.8,0,0}

\title{Probabilistic Mid- and Long-Term Electricity Price Forecasting}  

\author{\small Florian Ziel, Universit\"{a}t Duisburg-Essen, Berliner Platz 6-8, 45127 Essen, Germany \\
\small Rick Steinert, Europa-Universit\"{a}t Viadrina, Gro\ss e Scharrnstra\ss e 59, 15230 Frankfurt (Oder), Germany}

\begin{document}
\maketitle
\lhead{\nouppercase{\leftmark}}
\begin{abstract}
The liberalization of electricity markets and the development of renewable energy sources has led to new challenges for decision makers. These challenges are accompanied by an increasing uncertainty about future electricity price movements. The increasing amount of papers, which aim to model and predict electricity prices for a short period of time provided new opportunities for market participants. However, the electricity price literature seem to be very scarce on the issue of medium- to long-term price forecasting, which is mandatory for investment and political decisions. Our paper closes this gap by introducing a new approach to simulate electricity prices with hourly resolution for several months up to three years. Considering the uncertainty of future events we are able to provide probabilistic forecasts which are able to detect probabilities for price spikes even in the long-run. As market we decided to use the EPEX day-ahead electricity market for Germany and Austria. Our model extends the X-Model which mainly utilizes the sale and purchase curve for electricity day-ahead auctions. By applying our procedure we are able to give probabilities for the due to the EEG practical relevant event of six consecutive hours of negative prices. We find that using the supply and demand curve based model in the long-run yields realistic patterns for the time series of electricity prices and leads to promising results considering common error measures. 

\end{abstract}

\textbf{Keywords:} electricity prices; probabilistic forecasting; supply and demand; long-term; negative prices; renewable energy

\blfootnote{\textit{Abbreviations:}
ACE, average coverage error;  
AIC, Akaike information criterion;
AMAPE, adapted MAPE;
ANEM, Australian National Electricity Market;
ANN, artificial NN;
ANOVA, analysis of variance;
AWPI, average width of PIs;
ARMAX, autoregressive moving average model with exogenous inputs;
ARX, autoregressive model with exogenous inputs;
BNetzA, German Federal Network Agency;
BS, Brier Score;
CRPS, continuous ranked probability score;
CT, Christoffersen test;
CWC, coverage width-based criterion;
DC, direct current;
DWD, German Meteorological Office;
ECP, empirical coverage probability; 
ECR, evaluation criterion of resolution;
EEG, German Renewable Energy Sources Act;
ELM, extreme learning machine;
ENTSO-E, European Network of Transmission System Operators;
EPEX, European Power Exchange;
EUPHEMIA, Pan-European Hybrid Electricity Market Integration Algorithm;
GAMLSS, generalized additive models for location, scale and shape;
GARCH, generalized autoregressive conditional heteroskedasticity;
GDP, gross domestic product;
GME, Gestore dei Mercati Energetici;
k-NN, k nearest neighbor;
Lasso, least absolute shrinkage and selection operator;
LS, logarithmic scores;
LSSVM, least-squares SVM;
MAE, mean absolute error;
MAPE, mean absolute percentage error;
MLP, multi-layer perceptron;
MPIW, mean prediction interval width;
MSE, mean squared error;
MSPE, mean squared percentage error;
NN, neural network;
NLPD, average negative log predictive density;
NMPIW, normalised MPIW;
OLS, ordinary least squares;
OMIE, OMI-Polo Español;
PBS, pinball score/loss;
PCA, principal component analysis;
PCR, Price Coupling of Regions;
PI, prediction interval;
PICP, PI coverage probability; 
PINAW, PI normalised average width;
PITS, probability integral transform scores;
PJM, Pennsylvania-New Jersey-Maryland Interconnection;
RBF, radial basis function;
RE, reliability evaluation criterion;
RMSE, root MSE;
SARIMAX, seasonal autoregressive integrated moving average with exogenous inputs;
SE, sharpness criterion/score;
SVM, support vector machine;
UK, United Kingdom;
US, United States;
VEC, vector error correction model;
WNN, weighted nearest neighbor;
WS, Winkler score;
}


\section{Introduction {and Motivation}} \label{Introduction}
The past decades in electricity price research were characterized by the rapid liberalization process of several electricity markets across the world and the increasing development of renewable energy. Either voluntarily or by regulation, many institutions in the field of electricity contributed to an continuously improving transparency and quality of mostly freely available information on electricity prices and related time series. This in turn has helped researchers and practitioners to understand the mechanics of the price formation and lead to a large amount of papers which focus on electricity price forecasting. According to Weron, there was only a negligible amount of papers published before the year 2000, whereas in 2005 and 2006 the amount of papers reached their first peak point followed up by its hitherto maximum in 2009 \citep{weron2014electricity}. \par 
Research in electricity price forecasting originates from many different fields of science, e.g. engineering or statistics, which led to a manifold structure of different approaches. However, most of these approaches have in common that they focus on forecasting electricity prices in the short-term, specifically up to one day ahead with an hourly resolution (see e.g. \cite{aggarwal2009electricity} or \cite{weron2014electricity} for a literature review on electricity price forecasting). In contrast to this, electricity price forecasting methods which consider a longer period of time are rare \citep{yan2013mid}. A large proportion of research for that time horizon originates from fundamental models, which capture the dynamics of the system, e.g. the estimated cost functions of the market participants \citep{bello2016parametric,bello2016probabilistic}. These model types often lack to use realistic time series of prices and related data and therefore cannot provide a realistic hourly resolution of price predictions, which is 
typically the case in day-ahead markets. \par
Nevertheless, there are some models which are able to capture the hourly behavior of electricity price and provide mid- to long-term forecasts. Even though the literature is not consentaneous on this issue, we refer to the time horizon of one month to one year as mid-term and to the time horizon of more than one year as long-term. The model of \cite{barquinhybrid2008hybrid} for instance consists of a hybrid approach using fundamental and econometric, e.g. autoregressive, modeling techniques. They are able to utilize the hourly day-ahead electricity price series of Spain to forecast the whole year 2005. Yan and Chowdhurry were able to use data mining techniques, e.g. support vector machines (SVM), to study the PJM market in 2013 and 2015. In 2013 they show by a forecasting study that combining a least squares support vector machine with an ARMAX model yields promising results when the hourly forecast of one month is considered. For their setup they use training data of one year, e.g. 2009, to forecast the month 
of July 2010 \citep{yan2013mid}. Applying a two-stage SVM in 2015 they extend their model to be able to capture severe price peaks, which they describe as extremely difficult to model in a mid-term forecasting setting \citep{yan2015midterm}. As in their previous paper, they forecast one month with hourly resolution. 

Another important limitation of most electricity price forecasting models is their focus on specific moments of the distribution, particularly the mean and the variance. Also machine learning techniques commonly concentrate on point forecasts - which is a comparable counterpart to forecasting the mean in an econometric setting. Even though it seems most important at first glance to get point or mean forecasts for the electricity price, it is often the uncertainty of prices which has the highest impact for market participants. Achieving a precise point forecast may provide a solid basis for flexible investment decisions, but cannot account for the likelihood of possible extreme events, which can have tremendous consequences for the business as a whole. However, some researchers tackle this issue by analyzing and modeling the variance as well. But the concept of variance alone is not enough to quantify uncertainty in the case of electricity prices, as they usually tend to have non-symmetric heavy-tailed 
distributions which also vary over time. Hence, a possible solution for this issue can be to model the whole time-dependent distribution function of prices. 
{This field of research was considered in electricity price forecasting especially in recent years.} 
It can be summarized under the discipline of probabilistic forecasting.
One of the early papers covering probabilistic forecasts in terms of interval forecasts originated in 2006 by \cite{misiorek2006point}. They utilize well-known point forecasting models like ARX with GARCH components to construct interval forecasts for the hourly electricity price of the California Power Exchange. Later contributions, which explicitly focus on probabilistic forecasting emerged from econometric as well as machine learning approaches. For instance, Kou et al. were able to achieve day-ahead probabilistic forecasts for several electricity markets by combining machine learning with a variational heteroscedastic Gaussian process \citep{kou2015probabilistic}. A common approach during the recent years for econometric probabilistic forecasting was constructing prediction intervals by quantile regression. A basic introduction for this topic can be found in \cite{nowotarski2014computing}, among others. Extensions include for instance the Factor Quantile Regression Averaging of \citet{maciejowska2015probabilistic} or lasso-based approaches as done by \cite{gaillardsemi2015}.

{A recent review done by \cite{nowotarski2017recent} focused on the raising awareness of probabilistic forecasting in electricity price forecasting. They support their argument of an increased necessity of these methods by quantifying the development of published articles from the year 2003, where the first related article was published by \cite{zhang2003energy} and the year 2016. Given their numbers they show that probabilistic electricity price forecasting gained a tremendous increase in 2016, when the amount of published papers almost quadrupled from 3 in 2015 to 11 in the year 2016 by the time of their study.}

Another important direction of electricity price models originates from fundamental or structural electricity price models.
For these models the electricity price is considered as an equilibrium of supply and demand (see e.g. \cite{hirth2013market}, \cite{dillig2016impact}, \cite{pape2016fundamentals}). Here the major price drivers are the fundamental inputs like the load and the merit order curve and especially the marginal cost of the available power plant portfolio. These models are popular for long term forecasting of electricity prices as impacts of regulative changes, for example the closing of a certain power plant or a newly installed wind farm, can be easily drawn. However, these models usually only model the equilibrium price, e.g. the mean electricity price at a certain time point, but not the full underlying distribution. Even though new approaches in this direction like \cite{nahmmacher2016carpe} try to overcome this problem particularly by providing a representative fundamental market situation, the general problem that no temporal dependency information is used remains unsolved.

\begin{figure}[htb!]
 \includegraphics[width=1\textwidth]{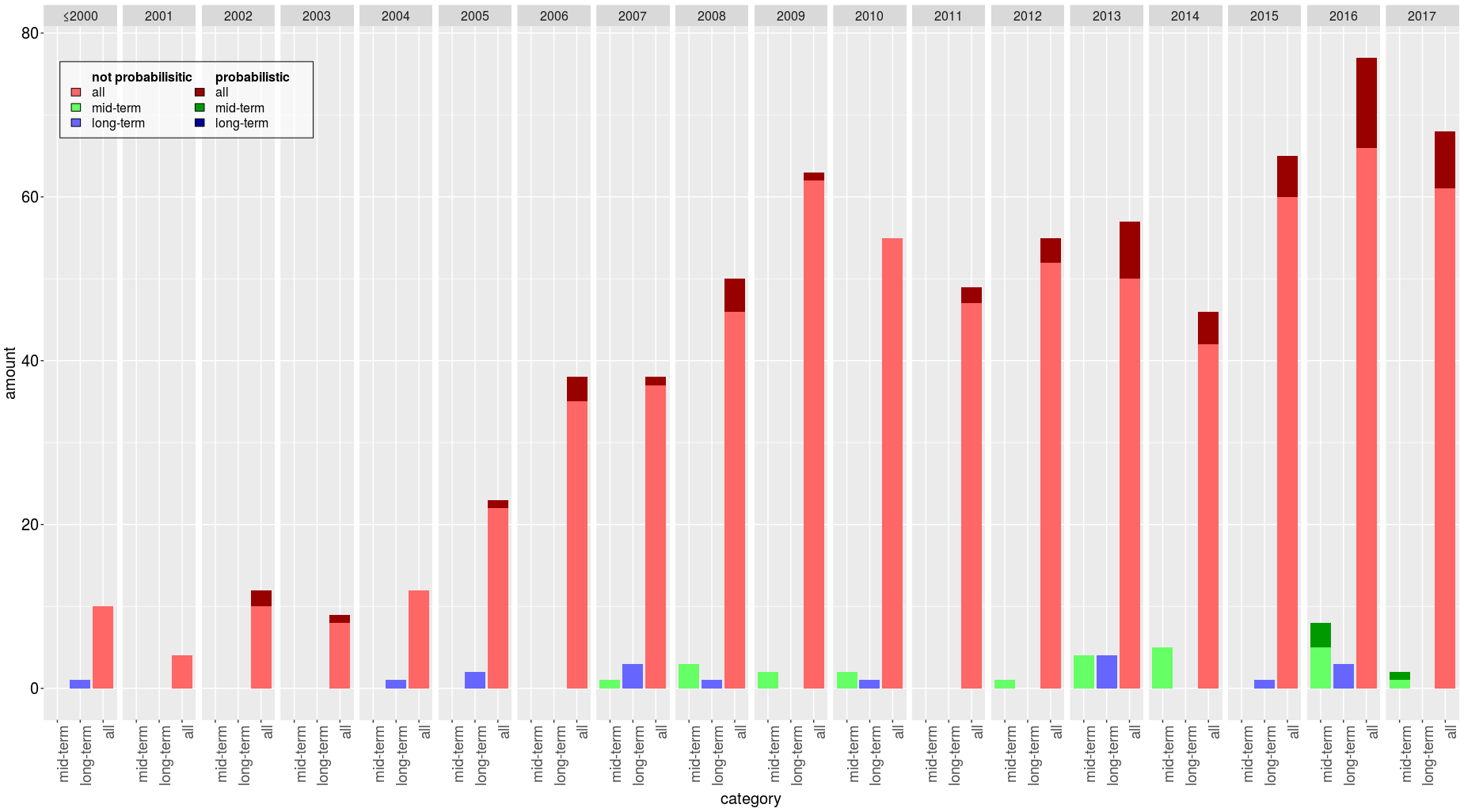}
\caption{{Amount of published papers up to 2017, categorized by forecasting horizon and whether the forecast was probabilistic or not.}}
\label{fig_review}
 \end{figure}

{Many researchers who conduct a review study seem to mainly focus on the difference in models and try to compare them by presenting overviews or their popularity over time (\cite{aggarwal2009electricity,nowotarski2017recent,
weron2014electricity}; among others). This focus resembles a point of view where the method is of utmost importance and not necessarily the purpose of modeling. However, from a practical standpoint it could be argued that the way how electricity prices are modeled is not as much important as the goal the modeling strategy actually pursues. For instance, if an electricity company is interested in building a new power plant, they will mainly be interested in long term electricity price forecasts over the whole lifetime of the plant, rather than to focus on a short-term horizon. Comparing different model strategies however may prove not to be too useful in this situation as it requires deep knowledge about properties and limitations of these models. Therefore we decided to present a brief review of the literature with the focus of the actual purpose of modeling rather than the model itself. This review part can therefore be considered as an update to e.g. \cite{nowotarski2017recent} with taking a new point of view on the issue. Hence, we will continue with a review of the recent and practical relevant topics of forecasting horizon and probabilistic forecasting. 

For our study we divided these two categories in the following sub-categories. Forecasting horizon was divided into mid-term and long-term forecasting as well as a category which accumulates over all horizons. This last category was chosen to easily determine the relative amount of mid- and long-term forecasting compared to all price forecasting horizons. Probabilistic forecasting is a binary category which acknowledges a paper as probabilistic forecasting whenever the full density of prices is forecasted. This specifically excludes papers which only forecast mean or variance. To analyze the amount of papers published in this field, we used Scopus as it is not only a well-known and reliable source for papers but also has an user-friendly interface for refined queries. To achieve rigour we geared the applied keywords towards the study of \cite{nowotarski2017recent} with minor changes. These keywords were combined by logical links to create the database-syntax specific queries for Scopus with the requirements for our categories. Hence, we needed to conduct six different queries, one for all electricity price forecasting papers,\footnote{{ (TITLE((((("electric*" OR "energy market" OR "power price" OR "power market" OR "power system" OR pool OR "market clearing" OR "energy clearing") AND (price OR prices OR pricing)) OR lmp OR "locational marginal price") AND (forecast OR forecasts OR forecasting OR prediction OR predicting OR predictability OR "predictive densit*")) OR ("price forecasting" AND "smart grid*")) OR TITLE-ABS("electricity price forecasting" OR "forecasting electricity price" OR "day-ahead price forecasting" OR "day-ahead mar* price forecasting" OR (gefcom2014 AND price) OR (("electricity market" OR "electric energy market") AND "price forecasting") OR ("electricity price" AND ("prediction interval" OR "interval forecast" OR "density forecast" OR "probabilistic forecast"))) AND NOT TITLE ("unit commitment")) AND (EXCLUDE(AU-ID,"[No Author ID found]" undefined))
}} one for mid-term forecasting,\footnote{{Query of footnote 1 combined with: AND ( TITLE("mid-term") OR TITLE("mid term") OR TITLE("medium-term") OR TITLE("medium term") OR TITLE("long-term") OR TITLE("long term") OR TITLE-ABS-KEY("mid-term electric*") OR TITLE-ABS-KEY("mid term electric*") OR TITLE-ABS-KEY("medium-term electric*") OR TITLE-ABS-KEY("medium term electric*") OR TITLE-ABS-KEY("long-term electric*") OR TITLE-ABS-KEY("long term electric*") OR TITLE-ABS-KEY("mid-term price*") OR TITLE-ABS-KEY("mid term price*") OR TITLE-ABS-KEY("medium- term price*") OR TITLE-ABS-KEY("medium term price*") OR TITLE-ABS-KEY("weeks-ahead") OR TITLE-ABS-KEY("weeks ahead") OR TITLE-ABS-KEY("month-ahead") OR TITLE-ABS-KEY("month ahead") OR TITLE-ABS-KEY("mid term* horizon") OR TITLE-ABS-KEY("mid-term* horizon") OR TITLE-ABS-KEY("medium term* horizon") OR TITLE-ABS-KEY("medium-term* horizon") )}} another one for long-term forecasting\footnote{{Query of footnote 1 combined with: AND ( TITLE("long-term") OR TITLE("long term") OR TITLE-ABS-KEY("long-term electric*") OR TITLE-ABS-KEY("long term electric*") OR TITLE-ABS-KEY("long-term price*") OR TITLE-ABS-KEY("long term price*") OR TITLE-ABS-KEY("year-ahead") OR TITLE-ABS-KEY("year ahead") OR TITLE-ABS-KEY("long term* horizon") OR TITLE-ABS-KEY("long-term* horizon") )}} and three queries for the before mentioned sets but with the restriction of only probabilistic {forecasting.\footnote{{Queries of footnotes 1 to 3 each combined with: AND (TITLE( ("probabilistic" AND "forecasting") OR interval OR density) OR TITLE-ABS-KEY("probabilistic forecast*" OR "interval forecast*" OR "density forecast*" OR "prediction interval*"))
}} After running each query we refined our search results by manually selecting only articles, books, book chapters, proceedings and editorials written in English. The results, ordered by our specific category subsets are presented in Figure \ref{fig_review}. 

This figure displays the amount of published papers in a specific category as bars, divided horizontally by the forecasting horizon. If a paper used probabilistic forecasting methods, the corresponding vertical area of the bar was dyed into a dark color of the same color as the associated forecasting horizon. This helps visually depicting the binary character of the usage of probabilistic forecasts. Using this color scheme it is easily obtainable that the overall amount of probabilistic forecasts compared to non-probabilistic forecasts is quite scarce. Only around 7.2\% of all 710 published papers we detected in our study were related to probabilistic forecasting. However, the amount of published papers with mid- or long-term forecasting was not much higher, resulting in a relative share of only 8\%. Given these two numbers it is of no surprise that the percentage of mid- or long-term forecasting papers with probabilistic forecasting is only 0.6\%. As can be concluded by examining Figure \ref{fig_review} these papers were only published in 2016 and 2017 respectively and only concerned mid-term forecasting. After investigating these papers in detail, we found that every of these papers was published by the same team of authors, with slightly changing co-authors \citep{bello2016parametric,bello2016medium,bello2016probabilistic,bello2017medium}.}
}

In order to grant a detailed insight into the most recent research for mid- and long-term probabilistic forecasting, we present the research articles of the seven most recent years (2011 to 2017) in tables \ref{tab_review},\ref{tab_review2} and \ref{tab_review3}.

\begin{savenotes}
\begin{table}[htb!]
\resizebox{\textwidth}{!}{%
\begin{tabular}{C{6.5cm}|C{4.5cm}|C{7.5cm}|C{6.5cm}|C{3.5cm}}
\multirow{2}{*}{\textbf{Reference}} & \multirow{2}{*}{\textbf{Market}} & \multirow{2}{*}{\textbf{Inputs}} & \multirow{2}{*}{\textbf{Model}}  & \textbf{Accuracy} \\ 
&&&&\textbf{Measures} \\ \hline
\multicolumn{5}{|c|}{} \\
\multicolumn{5}{|c|}{\textbf{Medium-term paper not probabilistic 2011-2017}} \\
\multicolumn{5}{|c|}{} \\ \hline

\multirow{3}{*}{\cite{yan2016performance}}&\multirow{3}{*}{PJM}&Electricity load, natural gas price,&\multirow{3}{*}{SVM and LSSVM}&\multirow{3}{*}{MAE, MAPE} \\
&&monthly averaged spot prices,&& \\
&&hour of the day and month&& \\ \hdashline

\multirow{3}{*}{\cite{mohamed2016mid}}&\multirow{3}{*}{New England}&Calendar day, fuel prices,  &\multirow{3}{*}{SVM}&\multirow{3}{*}{MAPE, RMSE, $R^2$} \\
&&electricity load, weather condition,&& \\
&&import and export&& \\ \hdashline

\multirow{3}{*}{\cite{cheng2016mid}}&\multirow{3}{*}{China Yunnan}&Electricity production, &\multirow{3}{*}{Grey prediction model} &\multirow{3}{*}{MAE, MAPE} \\
&&electricity load, export, && \\
&&number of generation/consumption companies&& \\ \hdashline
\multirow{3}{*}{\cite{chakravarty2016evolutionary}}&New England, Ontario, &\multirow{3}{*}{Spot prices}&Evolutionary-improved   &\multirow{3}{*}{MAPE, RMSE} \\
&PJM, California,&&cuckoo search& \\
&Nord Pool, Spain&&ELM& \\ \hdashline

\multirow{3}{*}{\cite{maciejowska2016short}}&\multirow{3}{*}{UK}&Spot prices, natural gas price,  &\multirow{3}{*}{AR, ARX, PCA}&\multirow{3}{*}{MAPE, RMSE} \\
&&electricity load, && \\
&&coal prices, CO2 prices && \\ \hdashline

\multirow{4}{*}{\cite{kossov2014medium}}&\multirow{4}{*}{25 countries}&GDP per capita,   oil spot price, &\multirow{4}{*}{Linear regression}&\multirow{4}{*}{MSPE}\\
&&national currency to USD, && \\
&&share of hydraulic and nuclear power, && \\
&&climate, year index\footnote{The author uses yearly data.} && \\ \hdashline

\multirow{3}{*}{\cite{yan2014mid}\footnote{The authors published several papers regarding the same market in 2013 and 2014, using an SVM in all articles.}}&\multirow{3}{*}{PJM}&Spot prices, electricity load, &\multirow{3}{*}{SVM}&\multirow{3}{*}{MAE, MAPE} \\
&&natural gas price,  price zones, && \\
&&month and hour of the day && \\ \hdashline

\multirow{3}{*}{\cite{voronin2014hybrid}}&\multirow{3}{*}{Nord Pool Finland}&Spot prices, electricity load,  &GMM,  &MSE,  \\
&&electricity supply, temperature, &k-NN, MLP,&MAE, \\
&&congestion, weekdays, holidays &SARIMAX-GARCH&AMAPE \\ \hdashline

\multirow{3}{*}{\cite{torbaghan2012medium}}&\multirow{3}{*}{Nord Pool, Ontario} & Spot prices, electricity load, & SVM,     & \multirow{3}{*}{MAPE} \\
&&temperature, hydro data, &RBF-NN,&\\
&&month index, holidays\footnote{The authors use monthly data.} &WNN& \\ \hline
\multicolumn{5}{|c|}{} \\
\multicolumn{5}{|c|}{\textbf{Medium-term paper probabilistic 2011-2017}} \\
\multicolumn{5}{|c|}{} \\ \hline
\multirow{5}{*}{\cite{bello2017medium}}&\multirow{5}{*}{Spain}&Spot prices, wind, hydro,    &  &  \\
&&import and export, load, net load,&Quadratic equilibrium model,&Percentage of  \\
&&  natural gas prices, coal prices,&Monte-Carlo,&Improvement, \\
&&CO2 prices, power plant unavailabilities,&Quantile regression,&WS, \\
&&power plant specific costs&Spatial interpolation&PBS \\ \hdashline

\multirow{5}{*}{\cite{bello2016medium}}&\multirow{5}{*}{Spain}&Spot prices, hydro, electricity load,   &Logistic regression, MLP     &   \\
&&wind, power plant specific costs &Monte-Carlo, Decision tree,& BS,\\ 
&& natural gas prices, CO2 prices, coal prices,&Spatial interpolation,& \\ 
&&power plant unavailabilities,& Markov Regime-Switching,&ECP \\ 
&&weekdays, holidays&Quadratic equilibrium model,& \\ \hdashline

\multirow{5}{*}{\cite{bello2016probabilistic}}&\multirow{5}{*}{Spain}&Spot prices, hydro, electricity load,  & &  \\
&&wind, power plant specific costs,  &Quadratic equilibrium model,&ECP, \\
&&natural gas
prices, CO2 prices, coal prices,&Monte-Carlo,& \\
&&power plant unavailabilities,&Spatial interpolation&PBS \\
&&weekdays, holidays&& \\ \hdashline

\multirow{5}{*}{\cite{bello2016parametric}}&\multirow{5}{*}{Spain}&Spot prices, wind, hydro,    & GAMLSS, &\multirow{5}{*}{ECP}  \\
&&import and export, load, net load,&Quadratic equilibrium model,&  \\
&&  fuel prices,&Monte-Carlo,& \\
&&power plant unavailabilities,&Quantile regression,& \\
&&power plant specific costs&Spatial interpolation& \\ \hline

\multicolumn{5}{|c|}{} \\
\multicolumn{5}{|c|}{\textbf{Long-term paper not probabilistic 2011-2017}} \\
\multicolumn{5}{|c|}{} \\ \hline
\multirow{3}{*}{\cite{de2016long}}&\multirow{3}{*}{Spain}&Spot prices, future prices,&\multirow{3}{*}{Cointegration, VEC}&\multirow{3}{*}{MAPE} \\
&&oil spot prices, oil future prices,&& \\
&&electricity load, wind generation&& \\ \hdashline
\multirow{3}{*}{\cite{kotur2016neural}}&\multirow{3}{*}{UK, Serbia}&Spot prices, electricity production,  &\multirow{3}{*}{MLP}&\multirow{3}{*}{MAPE} \\
&&electricity import and export,&& \\
&&seasonal and daytime indicators&& \\ \hdashline
\multirow{3}{*}{\cite{azadeh2013optimum}}&\multirow{3}{*}{Iran}&Electricity load,  &ANN,   ANOVA,  &\multirow{3}{*}{MAPE} \\
&&powerhouse efficiency, &Linear regression,& \\
&&inflation, fuel price\footnote{The authors use yearly data.} &Fuzzy linear regression,& \\ \hline
\multicolumn{5}{|c|}{} \\
\multicolumn{5}{|c|}{\textbf{Long-term paper probabilistic 2011-2017}} \\
\multicolumn{5}{|c|}{} \\ \hline
\multirow{5}{*}{This paper}&\multirow{5}{*}{Germany/Austria}&Spot prices, temperature, wind,&\multirow{5}{*}{Lasso, Bootstrap}& \\
&&solar, nuclear, lignite, coal&& \\ 
&&natural gas, water,&&ECP\\
&&electricity load, auction data,&&\\
&& day/week/season/holiday dummies&&\\ \hline
\end{tabular}
}
\caption{Detailed comparison of probabilistic and mid- to long-term electricity price articles published between 2011 and 2017.}
\label{tab_review}
\end{table}
\end{savenotes}

\pagebreak

\begin{savenotes}
\begin{table}[htb!]
\resizebox{\textwidth}{!}{%
\begin{tabular}{C{6.5cm}|C{4.5cm}|C{7.5cm}|C{6.5cm}|C{3.5cm}}
\multirow{2}{*}{\textbf{Reference}} & \multirow{2}{*}{\textbf{Market}} & \multirow{2}{*}{\textbf{Inputs}} & \multirow{2}{*}{\textbf{Model}} & 
\textbf{Accuracy} \\ 
&&&&\textbf{Measures} \\ \hline
\multicolumn{5}{|c|}{} \\
\multicolumn{5}{|c|}{\textbf{Probabilistic paper not medium- or long-term 2015-2017}} \\
\multicolumn{5}{|c|}{} \\ \hline

\multirow{3}{*}{\cite{rafiei2017probabilistic}}&&\multirow{3}{*}{Spot prices}& \multirow{1}{*}{ELM, Wavelet preprocessing} 
&\multirow{3}{*}{ECP, ACE, WS} \\ 
&Ontario, Australia ANEM&&Improved clonal selection algorithm& \\
&&&& \\ \hdashline

\multirow{3}{*}{\cite{tahmasebifar2017point}}&\multirow{3}{*}{New South Wales}&\multirow{3}{*}{Spot prices, electricity load}& \multirow{1}{*}{Wavelet
transformation} 
&\multirow{3}{*}{ACE, PINAW, WS} \\
&&&Bootstrap, Mutual Information & \\ 
&&&ELM& \\ \hdashline
\multirow{3}{*}{\cite{ji2017probabilistic}}&\multirow{3}{*}{none, simulated}&\multirow{1}{*}{Spot prices, electricity load}& \multirow{3}{*}{DC Optimal Power Flow} 
&\multirow{3}{*}{BS} \\
&&operation and contingency constraints&& \\
&&&& \\ \hdashline
\multirow{3}{*}{\cite{rafiei2017probabilistic2}}&\multirow{3}{*}{Ontario, Australia ANEM}&\multirow{3}{*}{Spot prices, electricity load}& \multirow{1}{*}{
Generalized ELM
} 
&\multirow{3}{*}{ACE, ECR} \\
&&&Wavelet neural networks,& \\
&&&Wavelet preprocessing, Bootstrap & \\ \hdashline

\multirow{3}{*}{\cite{wan2017pareto}}&\multirow{3}{*}{Victoria}&\multirow{3}{*}{Spot prices}& \multirow{1}{*}{ELM} 
&\multirow{3}{*}{ECP, ACE, AWPI} \\
&&&Pareto optimal & \\
&&&prediction interval construction& \\ \hdashline

\multirow{3}{*}{\cite{ziel2016electricity}}&\multirow{3}{*}{Germany/Austria}&\multirow{1}{*}{Spot prices, auction data,}& 
&\multirow{3}{*}{ECP} \\
&&electricity production, wind and solar &Lasso&   \\
&&&& \\ \hdashline

\multirow{3}{*}{\cite{moreira2016probabilistic}}&\multirow{3}{*}{Spain}&\multirow{1}{*}{Spot prices, 
electricity load}& \multirow{1}{*}{Hilbert Kernel Quantile regression}&\multirow{3}{*}{ACE, PINAW} \\
&&wind power, generation, wind speed,& Quantile boosting, Quantile ANN & 
\\
&&precipitation, temperature, solar irradiation && \\ \hdashline

%
&&\multirow{3}{*}{Spot prices}& \multirow{1}{*}{ELM} 
&\multirow{3}{*}{ECP, ACE, WS} \\
\cite{rafiei2016novel}& Ontario, Australia
ANEM&&Wavelet preprocessing, Bootstrap& \\
&&&& \\ \hdashline

\multirow{3}{*}{\cite{maciejowska2016probabilistic}}&\multirow{3}{*}{UK}&\multirow{3}{*}{Spot prices, electricity load}& \multirow{1}{*}{Factor quantile regression} 
&\multirow{1}{*}{ECP, WS} \\
&&&PCA, Forecast combination& CT \\
&&&& \\ \hdashline

\multirow{3}{*}{\cite{juban2016multiple}}&\multirow{3}{*}{US (GEFCom2014)}&\multirow{3}{*}{Spot prices, electricity load}
&\multirow{3}{*}{Quantile regression with $\|\cdot\|_2$-penalty}&\multirow{3}{*}{PBS} \\
&&&& \\
&&&& \\ \hdashline

\multirow{3}{*}{\cite{dudek2016multilayer}}&\multirow{3}{*}{US (GEFCom2014)}&\multirow{3}{*}{Spot prices, electricity load}& \multirow{3}{*}{Bayesian ANN} 
&\multirow{3}{*}{PBS} \\
&&&& \\
&&&& \\ \hdashline
\multirow{3}{*}{\cite{maciejowska2016hybrid}}&\multirow{3}{*}{US (GEFCom2014)}&\multirow{3}{*}{Spot prices, electricity load, holidays}& \multirow{1}{*}{
Pre-filtering, Forecast combination
} 
&\multirow{3}{*}{PBS} \\
&&&Linear regression, Quantile regression& \\
&&&Post-processing & \\ \hdashline

\multirow{3}{*}{\cite{gaillard2016additive}}&\multirow{3}{*}{US (GEFCom2014)}&\multirow{3}{*}{Spot prices, electricity load, holidays}& \multirow{1}{*}{Generalized additive models} 
&\multirow{3}{*}{PBS, ECP} \\
&&&Forecast combination&
 \\
&&&Kernel based quantile regression  & \\ \hdashline

\multirow{3}{*}{\cite{nowotarski2015computing}}&\multirow{3}{*}{PJM}&\multirow{3}{*}{Spot prices, temperature}& \multirow{1}{*}{Forecast combination, } 
&\multirow{1}{*}{ECP} \\
&&&Quantile regression& CT \\
&&&& \\ \hdashline

\multirow{3}{*}{\cite{shrivastava2015prediction}}&\multirow{3}{*}{Ontario}&\multirow{3}{*}{Spot prices}& \multirow{3}{*}{Support vector regression}
&\multirow{3}{*}{ECP, ACE, CWC} \\
&&&& \\
&&&& \\ \hdashline

\multirow{3}{*}{\cite{shrivastava2015prediction2}}&\multirow{3}{*}{Ontario}&\multirow{3}{*}{Spot prices, electricity load}& \multirow{1}{*}{Support vector regression} 
&\multirow{3}{*}{ECP, ACE, WS} \\
&&&with multi-objective optimisation& \\
&&&& \\ \hdashline

\multirow{3}{*}{\cite{kou2015probabilistic}}&\multirow{3}{*}{New South Wales}&\multirow{3}{*}{Spot prices, electricity load}& \multirow{1}{*}{Variational heteroscedastic} 
&\multirow{3}{*}{ACE, WS, NLPD} \\
&&&Gaussian process& \\
&&&with active learning& \\ \hdashline

\multirow{3}{*}{\cite{ji2015probabilistic}}&\multirow{3}{*}{none, simulated}&\multirow{1}{*}{Spot prices, electricity load}& \multirow{3}{*}{DC Optimal Power Flow} 
&\multirow{3}{*}{BS} \\
&&operation and contingency constraints&& \\
&&&& \\ \hdashline
\end{tabular}
}

\caption{Detailed comparison of probabilistic and mid- to long-term electricity price articles published between 2011 and 2017.}
\label{tab_review2}
\end{table}
\end{savenotes}

\begin{savenotes}
\begin{table}[htb!]
\resizebox{\textwidth}{!}{%
\begin{tabular}{C{6.5cm}|C{4.5cm}|C{7.5cm}|C{6.5cm}|C{3.5cm}}
\multirow{2}{*}{\textbf{Reference}} & \multirow{2}{*}{\textbf{Market}} & \multirow{2}{*}{\textbf{Inputs}} & \multirow{2}{*}{\textbf{Model}} & 
\textbf{Accuracy} \\ 
&&&&\textbf{Measures} \\ \hline
\multicolumn{5}{|c|}{} \\
\multicolumn{5}{|c|}{\textbf{Probabilistic paper not medium- or long-term 2011-2014}} \\
\multicolumn{5}{|c|}{} \\ \hline

\multirow{3}{*}{\cite{jonsson2014predictive}}&\multirow{3}{*}{Nordpool}&\multirow{3}{*}{Spot prices, electricity load and wind power}& \multirow{1}{*}{Conditional m-GARCH model,} 
&\multirow{3}{*}{CRPS} \\
&&&Quantile regression& \\
&&&& \\ \hdashline

\multirow{3}{*}{\cite{nowotarski2014merging}}&\multirow{3}{*}{Nordpool}&\multirow{3}{*}{Spot prices}& \multirow{1}{*}{Quantile regression,} & \multirow{3}{*}{ECP, CT} \\
&&&Forecast combination& \\
&&&& \\ \hdashline

&\multirow{3}{*}{Ontario}&\multirow{3}{*}{Spot prices, electricity load}& \multirow{1}{*}{Support vector regression} 
&ECP, PINAW, \\
 \cite{shrivastava2014prediction} &&&with multi-objective optimisation&CWC,  \\
&&&&WS \\ \hdashline

\multirow{3}{*}{\cite{wan2014hybrid}}&\multirow{3}{*}{New South Wales}&\multirow{3}{*}{Spot prices, electricity load}& \multirow{3}{*}{ELM, Bootstrap} & \multirow{3}{*}{ECP, ACE, WS} \\
&&&& \\
&&&& \\ \hdashline

\multirow{3}{*}{\cite{shrivastava2013point}}&\multirow{3}{*}{Ontario }&\multirow{3}{*}{Spot prices}& \multirow{3}{*}{ELM, Wavelets} & \multirow{3}{*}{ACE, PINAW, CWC} \\
&&&& \\
&&&& \\ \hdashline

\multirow{3}{*}{\cite{sharma2013hybrid}}&\multirow{1}{*}{Ontario, California}&\multirow{3}{*}{Spot prices}& \multirow{1}{*}{Recurrent Neural Network} & \multirow{3}{*}{ACE, CT} \\
&Victoria, && and
coupled excitable system& \\
&New South Wales, Spain&&& \\ \hdashline

\multirow{3}{*}{\cite{wu2013new}}&\multirow{3}{*}{New England}&\multirow{3}{*}{Spot prices}& \multirow{1}{*}{Recursive dynamic factor analysis,} & \multirow{3}{*}{ACE, WS} \\
&&&Kalman Filter& \\
&&&& \\ \hdashline

\multirow{3}{*}{\cite{khosravi2013neural}}&\multirow{3}{*}{Victoria, New York City}&\multirow{3}{*}{Spot prices}& \multirow{3}{*}{ANN, GARCH, Bootstrap} & \multirow{3}{*}{ECP, PINAW, CWC} \\
&&&& \\
&&&& \\ \hdashline

\multirow{3}{*}{\cite{khosravi2013quantifying}}&\multirow{3}{*}{Victoria, New York City}&\multirow{3}{*}{Spot prices}& \multirow{3}{*}{ANN} & \multirow{3}{*}{ECP, PINAW, CWC} \\
&&&& \\
&&&& \\ \hdashline

\multirow{3}{*}{\cite{klaeboe2015benchmarking}}&\multirow{3}{*}{Nordpool Norway}&Spot prices,  & \multirow{3}{*}{Markov model, SARMA} & \multirow{3}{*}{ECP} \\
&&balancing prices and volumes,&& \\
&&production volume&& \\ \hdashline

\multirow{3}{*}{\cite{ji2013forecasting}}&\multirow{3}{*}{none, simulated}&\multirow{3}{*}{Spot prices, electricity load}& \multirow{1}{*}{DC Optimal Power Flow,} & \multirow{3}{*}{-} \\
&&&Markov model& \\
&&&& \\ \hdashline

\multirow{3}{*}{\cite{huurman2012power}}&Nordpool Oslo,&\multirow{1}{*}{Spot prices,}& \multirow{3}{*}{ARIMAX-GARCHX} & \multirow{3}{*}{PITS, LS} \\
&Denmark East&temperature, precipitation and wind speed&& \\
&&&& \\ \hdashline

\multirow{3}{*}{\cite{bo2012probabilistic}}&\multirow{3}{*}{none, simulated}&\multirow{3}{*}{Spot prices, electricity load}& \multirow{1}{*}{Alternating current} & \multirow{3}{*}{-} \\
&&&optimal power flow& \\
&&&& \\ \hdashline

\multirow{3}{*}{\cite{chen2012electricity}}&\multirow{3}{*}{Queensland}&\multirow{3}{*}{Spot prices, electricity load}& \multirow{3}{*}{ELM, Bootstrap} & \multirow{3}{*}{ECP, ACE} \\
&&&& \\
&&&& \\ \hdashline

\multirow{3}{*}{\cite{serinaldi2011distributional}}&\multirow{1}{*}{California}&\multirow{3}{*}{Spot prices, electricity load}& \multirow{3}{*}{Generalized additive models} & \multirow{3}{*}{ECP} \\
&Italy&&& \\
&&&& \\ \hdashline

\multirow{3}{*}{\cite{alonso2011seasonal}}&\multirow{3}{*}{Spain}&\multirow{3}{*}{Spot prices}& \multirow{3}{*}{Seasonal dynamic factor analysis} & \multirow{3}{*}{ECP} \\
&&&& \\
&&&& \\ \hdashline


\end{tabular}
}

\caption{Detailed comparison of probabilistic and mid- to long-term electricity price articles published between 2011 and 2017.}
\label{tab_review3}
\end{table}
\end{savenotes}


The tables categorize these papers into the four possible categories for mid-term versus long-term forecasting in combination with probabilistic versus not probabilistic modeling respectively. It also contains the category of all forecasting papers focusing on probabilistic modeling in the short-term horizon. We compare the papers by their selected market, the chosen input factors, the utilized model and the used measures for the accuracy of their approach. We tried to standardize terms as much as possible and renamed them if they were at least very similar to each other. Examples include terms like electricity demand and load or relative frequency and empirical coverage which were harmonized to load and empirical coverage respectively.

Overall it can be seen that non-probabilistic modeling approaches in the mid- and long-run are quite homogeneous in their choice of accuracy measure. The vast majority measures quality by means of absolute deviation, mostly in term of MAE or MAPE. This finding is especially intriguing as most of these authors use quadratic optimization functions in lieu of absolute functions in order to estimate their models. Thus, most models forecast the mean where strictly proper evaluation (see \cite{gneiting2007strictly}) requires the (R)MSE instead of the MAE which is optimal for median forecasts. Furthermore, the MAPE is not a recommended evaluation criteria in general due to problems with zero and close to zero prices and resulting troubles in the significance analysis, see \cite{franses2016note}.
The model approaches among researchers in these two categories are quite diverse. However, two model types, the SVM and the linear regression most certainly stand out, even though they are often complemented by various other approaches. For the inputs a manifold amount of regressors have been introduced. Input factors like electricity load and specific calendar or weekdays are the most popular choices, the ladder especially due to their long-term deterministic structure. Providing reliable and authentic forecasts for electricity prices in the mid- and long-run requires the researcher to establish a model for most of the regressors as well, as most of them are random and therefore unknown in the future. We find that this in turn refrains some of the authors to chose highly correlated but yet hardly predictable regressors to be included into the model.

The overall picture for probabilistic models shows that for all categories authors pursue very different ideas. Even the measure for quality is tremendously diverse. It is obtainable that each author seems to have its favorite choice of measure and usually sticks to the choice, if several papers are published. The employed models also range from an enormous variety of approaches, e.g. machine learning, autoregression, quantile regression and simulation. Compared to not probabilistic modeling the set of input factors is usually sparse. This is very likely due to the higher intricacy of probabilistic approaches and the resulting issues of e.g. estimation/training time. Among the input factors besides historical prices, load data is most popular. Here the authors usually directly take load forecasts as an input or use a separate model for the load. 
The table also shows the beforementioned facts that there is only one group of authors so far working on medium-term probabilistic forecasts. In addition to that there is no manuscript, except for this article, which introduced a long-term probabilistic model.

Moreover, we also want to point out that the number of papers in the table does not necessarily match the number of papers shown in Figure \ref{fig_review}. Even though we used the same syntax and database, we manually decided to drop out some of the papers. The reason for that is twofold. 

Firstly, some of the papers were actually false flags, as they used our keywords inside their abstract but still did not forecast exchange electricity prices in the mid- or long-term. For instance \cite{nowotarski2016importance}, who only forecast the long-term seasonal component of prices, but not the electricity prices in the long-run themselves. Another example is \cite{raskin2015typical}, who focus on classification schemes for specific days for long-term forecasting but yet do not forecast the price.

Secondly, we decided to not list all papers of authors, if the published works were extremely similar to each other. This usually happens when authors publish their works at a conference and later in a modified version at a journal, see also footnote 6 for an example.

\begin{figure}[htb!]
 \includegraphics[width=1\textwidth]{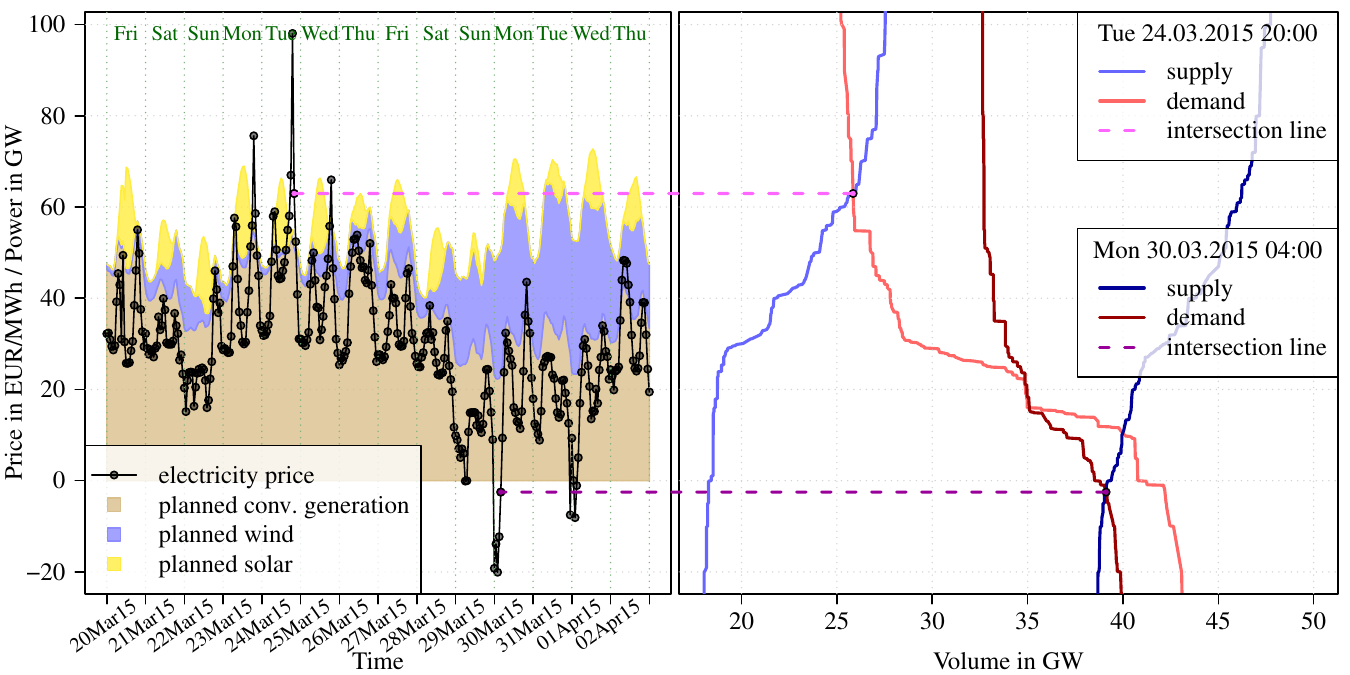}
\caption{Electricity prices with the corresponding planned generation by conventional power plants as well as wind and solar power (left) and the sale and purchase curves for two selected auctions and hours (right) with the corresponding intersection lines as equilibrium price.}
\label{fig_intro}
 \end{figure}

Summarizing these findings, we observe that electricity price forecasting approaches rarely focus on models, which are able to produce long-term forecasts with hourly resolution. By adding the important perspective of probabilistic forecasting there is no single paper known to us, which can account for all three components. Possible reasons for that are for instance  computational burdens due to high-dimensional datasets and too restrictive modeling perspectives. However, the recently developed X-Model, which considers the price as a function of supply and demand, provides a promising basis to meet all the requirements for probabilistic long-term modeling with hourly resolution. The model was proposed by \cite{ziel2016electricity} and focuses on the hourly sale and purchase curves for electricity, 
as illustrated in Figure \ref{fig_intro}. By analyzing the bidding behavior of market participants, the model is able to predict not only impending heavy price movements but also the whole time-dependent distribution function, including the stylized fact of price clusters. To construct forecasts the model utilizes the main source of price formation, e.g. the expectations of market participants about the key drivers of electricity prices. These expected or planned time series of conventional generation, wind and solar power are shown in Figure \ref{fig_intro}. Using these time series qualifies the model in general to incorporate possible shocks on the electricity market, e.g. the outage of power plants. Extending this model to a long-term forecast horizon will therefore also grant insights for the electricity price in long-term scenario analysis, when for instance the expansion of renewables is considered.

 Moreover, electricity producers or consumers need accurate forecasts for future events to reduce business risk. 
 For instance, renewable energy producers in Germany can get subsidies according to the German Renewable Energy Sources Act 2014 (EEG 2014).
 However, due to \S 24 of the EEG 2014 these subsidies are curtailed, if the electricity price is negative for at least six hours in a row. 
Those events where the electricity price is negative for six hours in a row are extremely difficulty to forecast, especially if they are far in the future.
One reason for that is that these events are usually very rare and therefore only a few observations to learn from are present. Moreover, to successfully predict such an event the dependency structure between the electricity prices
must be forecasted. Standard techniques that give an hourly price forward curve as well as so called probabilistic forecasting methods like quantile regression are not able to do so. 
The statistical reason behind it is that quantile regression only
allows the evaluation of the marginal distribution of the electricity price at a certain time point, but ignores the dependency structure that is relevant for events like the six consecutive hours of negative prices.
This paper presents a modeling methodology that allows for forecasting and estimating the probability of these events even far in the future.

The approach we will establish within this paper will therefore start with the basic X-Model and aims to amplify it to long-term forecasting. We will therefore elaborate a scheme which maps real observations of electricity production into market expectations. We will follow up with a rigorous presentation of the full density of up to a few years ahead forecasted electricity prices and present information on the important probability for six consecutive hours of negative prices. Hence, our paper is organized as follows. In section \ref{market-model} we will describe the neccessary model setup, the dataset and the market conditions of the EPEX SE for the auction of the day-ahead electricity spot price of Germany and Austria. Section \ref{model} provides a brief introduction to the X-model and will continue with a detailed description of the changes necessary to adjust the model to a long-term forecasting horizon. In the following we will describe our forecasting procedure briefly. The outcome of our forecasting study is presented in section \ref{results_discussion} which will also cover the results of the analysis for the likelihood of hours with negative prices. We will evaluate our approach by a common error measure for probabilistic forecasting and will provide a critical reflection of our work. Section \ref{conclusion} concludes our findings.

\section{Market Model} \label{market-model}

The model for the hourly German and Austrian EPEX day-ahead spot price is constructed in several steps.

\begin{figure}
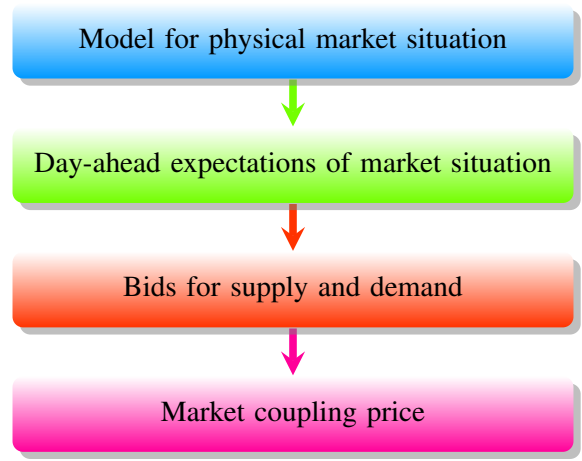

\begin{center}
\smartdiagramset{border color=none, text width=7.1cm, 
set color list={blue!40!cyan,green!40!lime,orange!40!red,red!40!magenta},
back arrow disabled=true}
\smartdiagram[flow diagram]{Model for physical market situation, Day-ahead expectations of market situation, Bids for supply and demand, Market coupling price}
\end{center}
\caption{Structure of modeling approach.}
\label{fig_model_flow}
\end{figure}

 Assume we want to estimate the electricity price of a day $d$.  
 As we model a day-ahead market were the auction is every day at 12:00 for the 24 hours (from 0:00 to 23:00) of 
 the next day we have to take into account that traders on the market must 
 make their bidding decision based on expectation for the next complete day. 
 Thus their bid relies heavily on the expectations of relevant input variables, such as the load or the wind power net feed-in 
 of the next day. These expectations are in {general} quite good, but still suffer from uncertainty. 
 Note that the predictions of many input variables used for making the bid decisions, such as wind power net feed-in, tend
 to have larger uncertainty for later hours. So the wind-power forecast for the 0:00 hour is usually
 much closer to the true value than the 23:00 hour. This is simply because of the fact that the latter event is further in the future
 and meteorologic models get worse precision with increasing forecast horizon. To tackle this issue we invented a specific model setup which uses four different steps. These steps, which are also illustrated in Figure \ref{fig_model_flow}, consist of:
\begin{enumerate}
 \item The physical market situation of a certain day, e.g. the actual load or wind power production, is simulated and considered as real observed time series.
 \item Based on the simulated physical market we construct the expected values
 of these time series, as if market participants would estimate the planned capacity for e.g. their power plants.
 \item Given the planned processes we utilize the expectations to construct the bidding behavior for the supply and the demand side of the market by using the X-Model.
 \item All the bids are aggregated and yield the sell and purchase curves of the market for which the intersection represents the market clearing price. 
\end{enumerate}
This model setup will be explained in detail in the following sections.

%
%
%

\subsection{The physical market situation model objects} \label{sec_subsec_phys_market_objects}

The physical market situation is described by a multivariate process of relevant hourly inputs.

These considered market information contains several processes.
{For our model setup they are the electricity generation of power plants of specific types, like nuclear, lignite, hard coal, natural gas and pump storage, where each energy source has its own time series. Moreover, we added the joint production of conventional power plants larger than 100 MW installed capacity as one single time series, the electricity consumption, the temperature and the wind and solar power net feed-in.} For the temperature we consider only the temperature of Frankfurt(Main) which is a city in Germany, relatively close to the center. 
Even though we consider the German/Austrian market, all these processes cover only German information due to a lack of availability of the considered data for Austria. Nonetheless, the electricity consumption of Austria is only about 10\% of the German electricity consumption. All production data originates from the EEX transparency database and the temperature data is downloaded from the Deutscher Wetterdienst (German Meteorological Office) climate data center.

\begin{figure}[htb!]
 \includegraphics[width=1\textwidth]{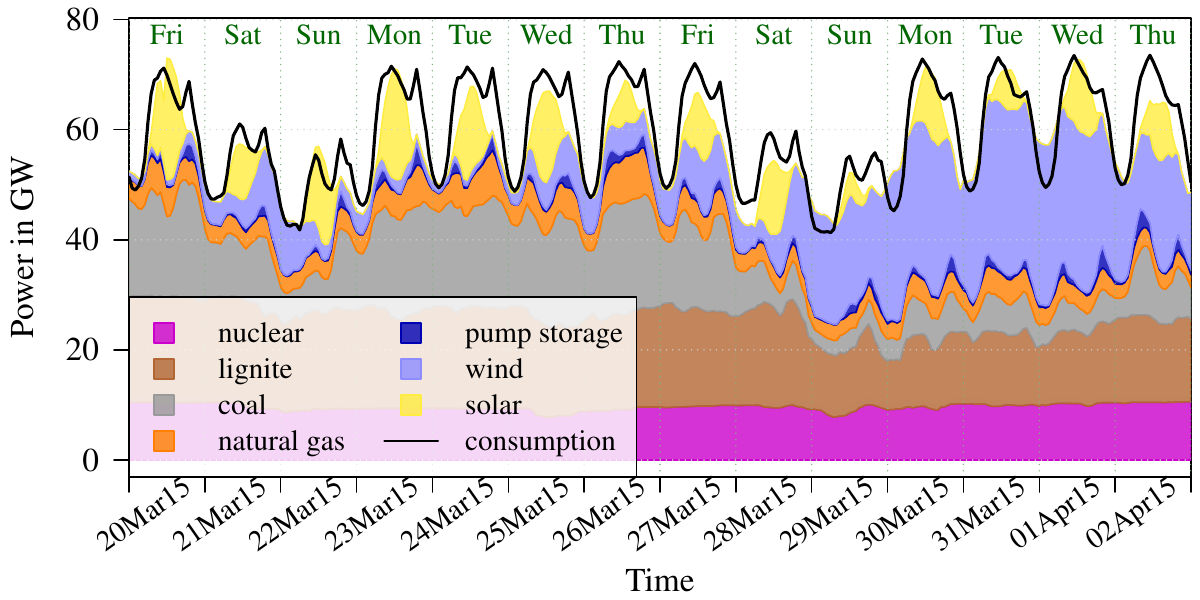}
\caption{{Snapshot of different time series which describe the relevant physical market situation from 20 March to 02 April 2015.} 
}
\label{fig_phy market}
 \end{figure}
In Figure \ref{fig_phy market} all the considered relevant processes, except for the temperature, are illustrated for the same period of time as used in Figure \ref{fig_intro}.
 Note that the considered power generation does not always sum up to the full consumption for two reasons. The first one is that 
 some electricity sources are ignored in the study, for example biomass and hydro power, mainly because of the data quality not being sufficient. The second reason is that 
 electricity imports and exports of electricity are ignored. As Germany is usually an electricity exporting country, it happens that the sum of the considered generation sources is larger than the consumption even though not all power sources are included.

For the considered processes we assume a specific relationship. For example the temperature influences the electricity consumption, but not the other way around.
Furthermore, we decided to adjust the wind and solar power feed-in by the installed capacity. The main reason for this is that we wanted to smooth these time series to increase predictability. This procedure also guarantees that we take the increased production over the years into account. We define the capacity adjusted wind and solar generation process to be the wind and solar net feed-in divided by the installed capacity of this time point. This process is more or less boiled down to the wind and sunshine effects. 
The installed capacity data originates from the Bundesnetzagentur (German Federal Network Agency).

As the processes evolve over time we consider an autoregressive model for $\bsY_t$ that is able to capture all relevant known stylized facts
, see e.g. \cite{paraschiv2014impact} or \cite{ziel2015efficient}.
These processes can be partitioned into two groups. 
The first one contains processes that are mainly driven by human impact, the other one 
are mainly driven by meteorologic influence. 
In our situation, only the temperature, the wind and solar processes fall into the meteorologic class.
All other inputs fall into the first class where we have clear human impact, especially weekend effects and public holiday effects.

\subsection{Day-ahead expectations on the physical market situation}
In order to determine the correct bid price, market participants have to make their expectations about all the before mentioned time series. Due to the day-ahead market structure, they have to do this for a time horizon of 24 hours of the next day. To determine a market price participants cannot know the physical market setting which corresponds to the day-ahead price they trade, as
this information concerns the future. It is therefore necessary, to transfer the physical market setting towards the market participants expectations, before we can directly implement them into the X-model.

For the $3$-dimensional process of real produced electricity $Y_{\text{gen.},t}, Y_{\text{wind},t}, Y_{\text{solar},t}$, for which $gen$ represents the generation of all conventional power plants, we observe the corresponding expectations for these time series one day in advance as planned production, as they are published by the EEX transparency. Hence, we denote the corresponding planned or day-ahead expected processes of $Y_{\text{gen.},t}, Y_{\text{wind},t}, Y_{\text{solar},t}$ by $Y_{\text{exp. gen.},t}, Y_{\text{exp. wind},t}, Y_{\text{exp. solar},t}$ as expectation for the physical market situation. To extend the notation from $t$ to a more detailed day and hour representation we will use $Y_{\text{exp. gen.},d,h}, Y_{\text{exp. wind},d,h}, Y_{\text{exp. solar},d,h}$, where $d$ is the day of time $t$ and $h$ is the corresponding hour of the day. As we can match the real production time series from the physical market setting with the expectations represented by the planned production we can learn from the decision making process over time and create our own expectation generating process using the physical market setting. The exact way this is done is described in section \ref{model}.

\subsection{Market model objects}
To map the expectations into the actual electricity price, we make use of the X-model of \cite{ziel2016electricity}. To utilize this model, we need to analyze and model the sale and purchase curves of the market.

The market clearing price is the result of an auction that takes place every day at 12:00 for the next day. 
For the German and Austrian EPEX spot price the EUPHEMIA algorithm is used since February 2014. This is 
a very complex algorithm developed from markets participating in the Price Coupling of Regions Initiative (PCR) like 
EPEX SPOT, Nord Pool, GME or OMIE. We use a simple but efficient approximation, that models the 
bids on the supply and demand side. By market regulation the price is defined to be between -500 and 3000 EUR/MWh.
Market participants can give their bid in this price span, where the minimal volume unit is 0.1 MWh and minimum price difference is 0.1 EUR/MWh. There 
are several different order products available, for example simple orders, block orders and smart orders, which are not included into our model setting. We approximate the market 
by regarding every bid as a simple order. Therefore we take the reported aggregated sale and purchase curve of the EPEX,
and compute the bid volume at each price and each time point as if there were no block products.
For a minimum price difference of 0.1 EUR/MWh on a possible price range of -500 to 3000 EUR/MWh this leads to 35001 possible volumes on 
the given price grid. To reduce the dimension we use the grouping approach from \cite{ziel2016electricity}. Here 
we create groups for the bidding behavior at the supply and demand side such that every 
bidding group contains in average about 1000 MWh of the bid volume.
So e.g. there could be a group ranging from 10.0 EUR/MWh to 15.3 EUR/MWh that contains all bid on the demand side. Note that for the 
supply side this usually results in a different restriction.

Every group is now considered a single time-series which is dependent on the expectations of the market participants regarding the distinct energy sources. Please note that the physical market situation therefore only has an indirect impact on the model, as we only use the expectations transitioned from the physical market and not the physical time series itself. From the different groups we are able to approximately reconstruct the original sale and purchase curves. For the exact description on how this is done, we refer to \cite{ziel2016electricity}.

\section{The time series models} \label{model}

As mentioned, we can divide the process into two classes. The physical market, that evolves over time and the bidding processes which consists of the expectations for the market situation one day ahead.
We usually have 24 hourly prices every day, except for the last Sunday in March, where only 23 hours are traded and the last Sunday in October, where 25 values are traded. As we want to exclude this effect, which is induced by the clock change we interpolate the missing hour in March and average the two 2 a.m. hours in October.

\subsection{Model for physical market situation} \label{model_physical}

{For the processes of the first class we consider a model that captures the autoregressive impact, 
the daily, weekly and annual seasonal behavior and public holidays effects as well as interaction effects.} The general construction of the models follows mainly the probabilistic 
load and temperature forecasting model of \cite{ziel2016lasso}. It 
proved high forecasting accuracy in forecasting competitions for short and long term forecasting, see \cite{hong2016probabilistic}.

In general, for all processes the assumed model is given by
\begin{equation}
 Y_{i,t} = \sum_{ l\in \II_i} \psi_{i,k} U_{k,t}  + \sum_{j=1}^D \sum_{k\in \JJ_{i,j} } \phi_{i,j,k} Y_{j,t-k} + \eps_{i,t}
 \label{eq_1dim}
\end{equation}
for each $i\in \{1,\ldots, D\}$ where $\psi_{i,k}$ and $\phi_{i,j,k}$ are parameters, $U_{k,t}$ are external regressors and $\eps_{i,t}$ 
the error term. The first sum contains dummy information, such as
seasonal cycles or public holiday impacts. In detail, the index sets $\II_i$ describe the active dummy information. The index sets $\JJ_{i,j}$
specify the autoregressive dependency structure. So it gives the lags of time series $j$ that have a potential impact on the time series $i$. 

\begin{figure}
 \centering

 \tikzstyle{block1} = [rectangle, draw, fill=blue!20, text centered, rounded corners, minimum height=1.8em]
\tikzstyle{block2} = [rectangle, draw, fill=blue!20, text centered, rounded corners, minimum height=1.1em]
    
\tikzstyle{line} = [draw, -latex']
    
\begin{tikzpicture}[node distance = 2cm, auto]
    \filldraw[fill=lime!30!white, draw=black] (0.5,0.5) rectangle (8,3.6);
    \filldraw[fill=gray!10!white, draw=black] (8.1,-4.1) rectangle (0.4,0.1);
    \node [block1, fill=cyan!30] (cawind) at (3,3) {capacity adjusted wind};
    \node [block1, fill=yellow!30] (casolar) at (3,2) {capacity adjusted solar};
    \node [block1, fill=red!30] (temperature) at (3,1) {temperature};
    \node [block1, fill=yellow!40] (solar) at (6.8,1) {solar};
    \node [block1, fill=cyan!40] (wind) at (5.3,1) {wind};
    \draw [->, thick, dashed, to path={-| (\tikztotarget)}] (cawind) edge (wind);
    \draw [->, thick, dashed, to path={-| (\tikztotarget)}] (casolar) edge (solar);
    \node[draw=none, text width=5em] at (6.7,3) {\textit{meteorologic components}};
    \node [block1, fill=green!30] (consumption) at (2.3,-.5) {load/consumption};
    \draw [->, thick, to path={-| (\tikztotarget)}] (wind) -- (5.3,0.1);
    \draw [->, thick, to path={-| (\tikztotarget)}] (solar) -- (6.8,0.1);
    \draw [->, thick, to path={-| (\tikztotarget)}] (temperature) -- (3,0.1);
    \filldraw[fill=gray!20!white, draw=black] (8,-4) rectangle (4.5,0) node[anchor=north west] (gnw){};
    \draw [->, thick, to path={-| (\tikztotarget)}] (consumption) -- (4.5, -.5);
    \node[draw=none, text width=8em] at (6.4,-.4) {\textit{generation (fuels)}};
    \node [block2, fill=magenta!40, anchor=west] (nuclear) at (4.75,-1) {nuclear};
    \node [block2, fill=brown!40, anchor=west] (lignite) at (4.75,-1.6) {lignite};
    \node [block2, fill=black!40, anchor=west] (coal) at (4.75,-2.2) {coal};
    \node [block2, fill=orange!40, anchor=west] (natural gas) at (4.75,-2.8) {natural gas};
    \node [block2, fill=blue!40, anchor=west] (pumpstorage) at (4.75,-3.4) {pump storage};
 
    \node [block1, fill=olive!40, text width=6em] (convgen) at (2.3,-1.8) {conventional generation};
    \draw [->, thick, to path={-| (\tikztotarget)}] (consumption) -- (convgen);
    \draw [->, thick, to path={-| (\tikztotarget)}] (4.5, -1.8) -- (convgen);
    \node[draw=none, text width=8em] at (2.4,-3.3) {\textit{consumption and production based components}};
 
\end{tikzpicture}
\caption{Dependency structure of the physical market situation components. The solid arrows represent a causal autoregressive relationship, the dashed arrows represent 
a functional relationship due to a capacity adjustment.}
\label{fig_phys_market}
\end{figure}
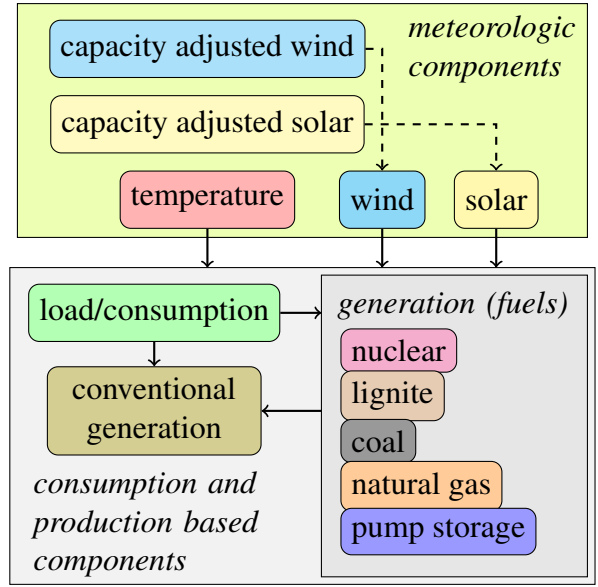

Furthermore, for each time series we distinguish between two different types of modeling:
If $Y$ depends on $X$ such that $Y_t$ depends on $X_{t-1}, X_{t-2}, \ldots$ we call this dependency \textit{autoregressive}. If $Y$ depends on $X$ such that $Y_t$ depends on $X_{t}, X_{t-1}, X_{t-2}, \ldots$, e.g. it uses the actual value of $X_t$, we call the dependency \textit{causal autoregressive}.
The dependency concerning the causally autoregressive characterization of the involved processes of the market situation is illustrated and given in Figure \eqref{fig_phys_market}.
The black solid arrow characterizes a causal autoregressive relationship. All components within a box of Figure \eqref{fig_phys_market} depend on each other autoregressively, e.g. the meteorologic components and the consumption/production based components. Note that the meteorologic components do not depend on the consumption/production time series, so for instance the temperature does not depend on the electricity consumption but obviously the consumption depends on the temperature. The dashed arrows illustrates the capacity transformation of the wind and solor power, which we receive easily by multiplying the seasonal adjusted wind and solar processes with the installed capacity at time $t$.
In detail, if a relationship between $i$ and $j$ in model equation \eqref{eq_1dim} is autoregressive we have $\JJ_{i,j} = \{1, \ldots, 360\}$ 
and if it is causally autoregressive we use $\JJ_{i,j} = \{0, \ldots, 360\}$. Hence, we allow for a possible memory of $360$ historic hours which corresponds to 15 days.

%
%


The external regressors $U_{k,t}$ are made of several components as similarly used in \cite{ziel2016lasso}. The full matrix $\bsU_t$ is given by
\begin{flalign}
\nonumber
\bsU_{t} =& ( 
(U^{\text{daily}}_{k,t})_{k\in \I_{\text{daily}}} , 
(U^{\text{weekly}}_{k,t})_{k\in \I_{\text{weekly}}} , 
(U^{\text{annual}}_{k,t})_{k\in \I_{\text{annual}}} ,\\
&(U^{\text{sm. annual}}_{k,t})_{k\in \I_{\text{sm. annual}}} ,
(U^{\text{d:sa}}_{k,t})_{k\in \I_{\text{d:sa}}} ,
(U^{\text{ph-fix}}_{k,t})_{k\in \I_{\text{ph-fix}}} ,
(U^{\text{ph-vary}}_{k,t})_{k\in \I_{\text{ph-vary}}} 
)
\end{flalign}
where we have daily, weekly, annual, smooth annual, daily-smooth annual interaction, fixed date holidays and varying date holidays effects.
For the meteorologic components we only consider daily, smooth annual and daily-annual interaction effects, as the 
other ones are clearly human driven due to week or holiday pattern.

The daily and weekly components are defined by
$ U^{\text{daily}}_{k,t} = k \leq \DD^{\text{daily}}(k) $
where $\I_{\text{daily} } = \{0, \ldots, 23\}$ and
$ U^{\text{weekly}}_{k,t} = k \leq \DD^{\text{weekly}}(k) $
where $\I_{\text{weekly} }= \{0, \ldots, 167\}$
and
$ U^{\text{annual}}_{k,t} = k \leq \DD^{\text{annual}}(k) $
where $\I_{\text{annual}}= \{0, \ldots, 365\}$.
The functions $\DD^{\text{daily}}(k)$,  $\DD^{\text{weekly}}(k)$,  $\DD^{\text{annual}}(k)$ give the hour of the day
($0, \ldots, 23$) and week ($0, \ldots, 167$) and the day of the year ($1, \ldots, 365$) of hour $k$ where the 29 February is considered 
as 28 February (both are considered as day 59 of the year). For the smooth annual basis { $U^{\text{sm. annual}}_{k,t}$ with corresponding index set $\I_{\text{sm. annual}}$}
we consider simply sine and cosine functions with a period of $365.24$ and $365.24/2$.
The interactions are defined by a multiplication of each daily component of $U^{\text{daily}}_{k,t}$
with the smooth annual component $U^{\text{sm. annual}}_{k,t}$.
This multiplication {term is denoted by $U^{\text{d:sa}}_{k,t}$ with corresponding index set $\I_{\text{d:sa}}$ and} allows for changes of the daily pattern over the year. This is usually very
distinct for meteorologic components, as the length of the night is shorter in summer in the northern celestial sphere than in winter.

The most complex external regressors concern the public holidays. 
Here we distinguish between fixed date and varying date public holidays, as considered in \cite{hong2010short} or \cite{ziel2016lasso}{, and define 
the holiday dummies $U^{\text{ph-fix}}_{k,t}$ and $U^{\text{ph-vary}}_{k,t}$ with index sets $\I_{\text{ph-fix}}$ and $\I_{\text{ph-vary}}$}.
Fixed date public holidays are every year at the same date, so e.g. 1 January (New Years Day) or 25 December (Christmas). Here the 
weekday when the public holiday occurs changes from year to year. 
Therefore the human impact is likely to change as well from year to year, thus the modeling is quite complicated.
For the dates with varying  holidays the date changes over the year. In Germany this affects Easter related holidays.
Still, here the weekday of the public holiday is fixed, e.g. Black Friday is on a Friday or Easter Monday is on Monday.
Hence, it makes sense to assume that the effects are similar every year. A more detailed definition of the used external regressors $U_{k,t}$ is presented in \cite{ziel2016lasso}.
As public holidays we consider all official public holidays of Germany, such as the important regional public holidays Corpus Christi, Epiphany, All Saints as well as Christmas Eve and New Years day.

\subsection{Model for day-ahead expectations of the physical market and bid processes}

As we are aiming to forecast the electricity price up to three years by forecasting the different bid price groups we need to forecast the expectations for the physical market also for up to three years. As this cannot be done by simply estimating and forecasting these time series of expectations with 365 or 366 days ahead, we had to create a specific update scheme for the expectations. 

\begin{figure}[htb!]
 \includegraphics[width=1\textwidth]{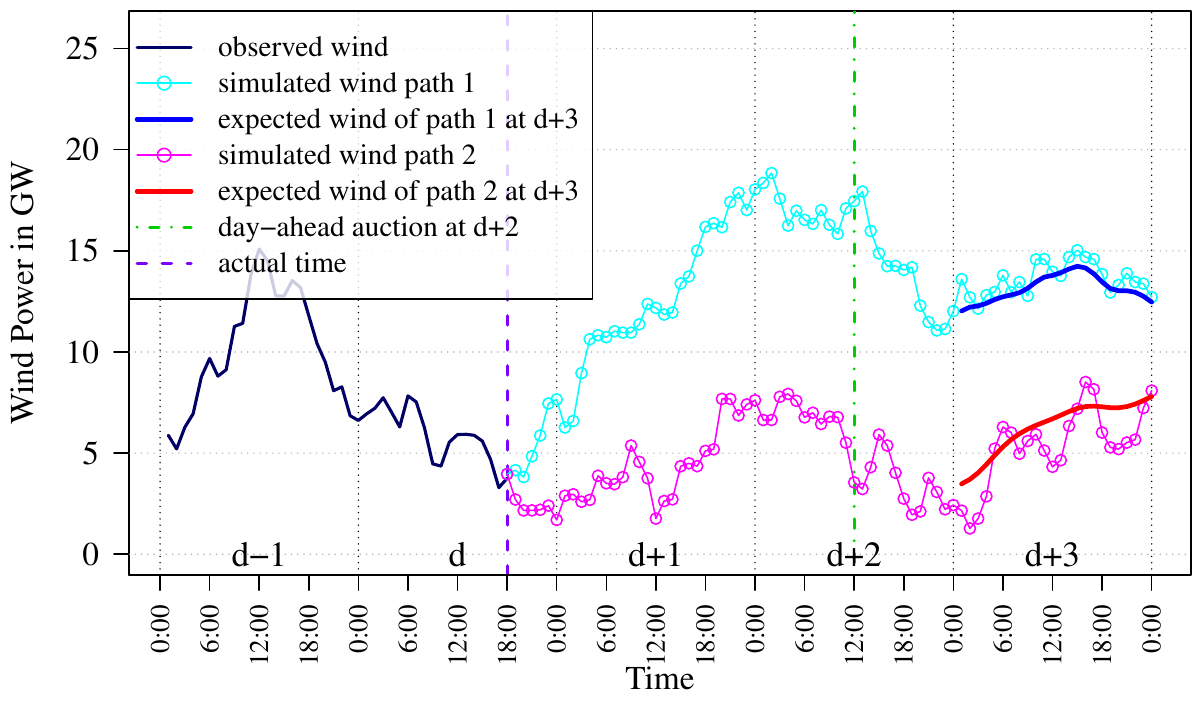}
\caption{{Different simulated wind production paths for a three days ahead forecast. The solid bold print lines in red and blue show the forecasted time series of expectations of the market participants given the simulated wind path right before the auction for the third day starts. These expectations are formulated on the assumption that these simulated paths were the real observable wind production.}}
\label{fig_exp_wind_example}
 \end{figure}

This procedure is illustrated as an example in Figure \ref{fig_exp_wind_example}. 
 There we assume that the current time is at day $d$ and $h=18$ and we are interested in a three days ahead forecast, so $d+3$.
For the day $d+3$ the auction takes place at $d+2$ and $h=12$.
At day $d$ 18:00 we can create a model for the physical market components as described in section  \ref{model_physical}. Given such a model we can simulate many wind {power production paths} 
power productions paths, which characterize the space of possible market situation outcomes. All these simulation paths are equally likely. In the figure there are two possible paths for the wind power generation given. 
A market participant could have observed e.g. the path 1 in cyan. On auction day $d+2$ at 12:00 the market participants have to make up their expectations on the market situation for all 24 hours of the next day $d+3$. These expectations are conditional expectations of all available data up to the day $d+2$ at $h=12$ where the day-ahead auction takes place. A real market participant would now make a forecast for the planned production based on the available information set up to that day and hour. However, as we already stated, the time series of market participants expectations was modeled so that it incorporates the real production value of the future. Given the simulation path 1 for the wind production we can now use this information additionally to create a smooth day-ahead expectation time series of the market participants, simply by treating this simulated path as real future path of wind production. Please note, that these expectations are not necessarily unbiased. The same explanation applies to the second simulated wind path in pink, which seems to shift in a slightly different direction. 


 Given the true market situation we can infer the expectations from the actual market situation. Therefore we use 
the available planned process $\bsZ_{t}$ which is the planned process of $\bsY_t$. $\bsZ_{t}$ is assumed to be available
to the market participants one day in advance.
Remember that the auction is every day at 12:00 and the day-ahead forecasts $\bsZ_{t}$ are available
for the full next day at this time point. Similarly we have the grouped bids as a 24-dimensional process.
Therefore we model these day-ahead expected processes by the 24 individual submodels. The general modeling approach is similar to \cite{ziel2016electricity}.

{Let $\bsZ_{d,h}$ denote the planned process at day $d$ and hour $h$, we can model $Z_{i,d,h}$ for every time series $i$ by
\begin{equation}
Z_{i,d,h} = \sum_{ k\in \II_{i,h}} \psi_{i,h,k} V_{k,d}  
+ \sum_{j=1}^D \sum_{l=1}^S \sum_{k\in \JJ_{i,j,l} } \phi_{i,j,l,k} Y_{j,d-k,l} 
+ 
\sum_{k\in \KK_{i,j,l} } \varphi_{i,j,l,k} Z_{j,d-k,l} 
+ \eps_{i,d,h}
\label{eq_24dim}
\end{equation}
Similarly as for the physical market situation we consider a specific autoregressive relationship.
So the planned processes of wind  and solar generation depend \textit{only} on the true values of the future, so that the 
forecasting behavior of the planned processes is well mapped. {This helps the model to learn from the expectation formation process given the true production process. When the out-of-sample forecasting is done for the physical market we are able to map the expectations from the forecast values as if they were the true values of the production process.} In contrast, the bidding groups depend on the physical market expectations, e.g. the day-ahead expectations of wind, solar and conventional generation, in a causal autoregressive way.
Additionally {these day-ahead market bids} also depend in a standard autoregressive way on their own history.

In detail we assume for a autoregressive relationship between $i$ at hour $h$ and $j$ at hour $l$ that
$$\JJ_{i,j,h} =\KK_{i,j,h} = \begin{cases}
    \{0,1, \ldots, 36 \}            & , i = j \text{ and } h=l \\
    \{0,1, \ldots, 8 \}            & ,   (i \neq j \text{ and } h=l ) \text{ or } (i = j \text{ and } h \neq l) \\
    \{0,1 \}            & ,   i \neq j \text{ and } h \neq l 
\end{cases}$$ 
and the autoregressive relationship
$$\JJ_{i,j,h} = \KK_{i,j,h} = \begin{cases}
    \{1, \ldots, 36 \}            & , i = j \text{ and } h=l \\
    \{1, \ldots, 8 \}            & ,   (i \neq j \text{ and } h=l ) \text{ or } (i = j \text{ and } h \neq l) \\
    \{1 \}            & ,   i \neq j \text{ and } h \neq l 
\end{cases}.$$ 
Hence, the potential memory of e.g. the supply bids at -500 EUR/MWh at day $d$ and hour $h=2$ can depend on the past 35 days of the bids at  -500 EUR/MWh at hour $h=2$. Additionally it can depend on up to the past 8 days of 
the bids at -500 EUR/MWh at other hours $j$ (so $j\neq h=2$) and the past 8 days of supply or demand bids at other price groups at hour $h=2$.
Finally, we allow that a bid  at  -500 EUR/MWh at hour $h=2$ can depend on all other bids of an arbitrary hour of the previous day.
This complex specification allows us to capture the most relevant dependency structure between the bids.
The external regressors $V_{i,h}$ contain as in \cite{ziel2016electricity} only week-day dummies which correspond to $U^{\text{weekly}}_{k,t}$ from the previous section.


\section{Estimation and Forecasting}

We conducted an estimation and forecasting study based on data from 1. November 2012 to 19. April 2015. 
We estimate the physical market model \eqref{eq_1dim}, the day-ahead expectations on the market model \eqref{eq_24dim}, and the electricity market bids \eqref{eq_24dim} using lasso.
The lasso estimation techniques introduced by \cite{tibshirani1996regression} can handle highly parameterized linear models efficiently using the coordinate descent 
estimation algorithm of \cite{friedman2007pathwise}.
For a linear model given by $Y = \bsbeta'\bsX + \eps$ where the elements of $\bsX$ have the same variance the lasso estimator is given by
\begin{equation}
 \what{\bsbeta} = \argmin_{\bsbeta} \|Y-\bsbeta'\bsX\|^2_2 + \lambda \|\bsbeta\|_1
\end{equation}
where $\lambda\geq 0$ is a tuning parameter and $\|\cdot\|_1$ and $\|\cdot\|_2$ are the standard $L^1$ and $L^2$ norms.
In this study we chose the optimal tuning parameter by minimizing the  Bayesian Information Criterion (BIC) as in \cite{ziel2016lasso} or \cite{ziel2016electricity}.

Given the data we perform a three year ahead ($3\times 365=1095$ days) forecasting study. The forecast is performed using residuals based bootstrap as in \cite{ziel2016lasso}. The residuals are sampled from all daily 24-dimensional residual vectors of all relevant processes\footnote{The univariate residual processes are transformed to a daily 24-dimensional process first}. Using the autoregressive formulation of all relevant time series (see \eqref{eq_1dim} and \eqref{eq_24dim}) we can easily simulate sample paths for the physical
market situation, the corresponding day-ahead  expectations as well as the supply and demand bids. Given the simulated supply and demand bids of a simulated day we compute the forecasted market clearing price for a specific day of a simulation path
by calculating the intersection of the supply and demand curves. In total we are using $N=10000$ sample paths for forecasting.

For computing the wind and solar power after simulating the capacity adjusted wind and solar power time series we require an assumption on the newly installed capacity. 
For the simulation study we considered the governmental plan for new installed capacities at the time when the forecast started. In detail the assumptions are based on the German Renewable Energy Sources Act from 2014 (EEG 2014). 
Within that Act, the installed solar power was planned to be increased by 2.5-3.5GWh annually, so we considered a linear annual growth of 3GWh. 
Similarly, for the wind power the plan was to annually increase the capacity by 2.4GWh to 2.6GWh on-shore and 0.75 GWh off-shore. Hence, we assume a linear wind power capacity growth of 3.25GWh per year.

\section{Empirical Results and Discussion} \label{results_discussion}

The model design allows us to estimate probabilities of basically all possible events or distributional characteristics. 
For probabilities of certain events, the relative frequencies in the $N$ sample paths can be used. 
For distributional characteristics like moments or quantiles we can use their sample estimates of the $N$ simulated paths to receive a suitable estimator. For illustration purpose we present some interesting results given the $N=10000$ simulated paths.

\begin{figure}[htb!]
\begin{subfigure}[b]{.99\textwidth}
 \includegraphics[width=1\textwidth]{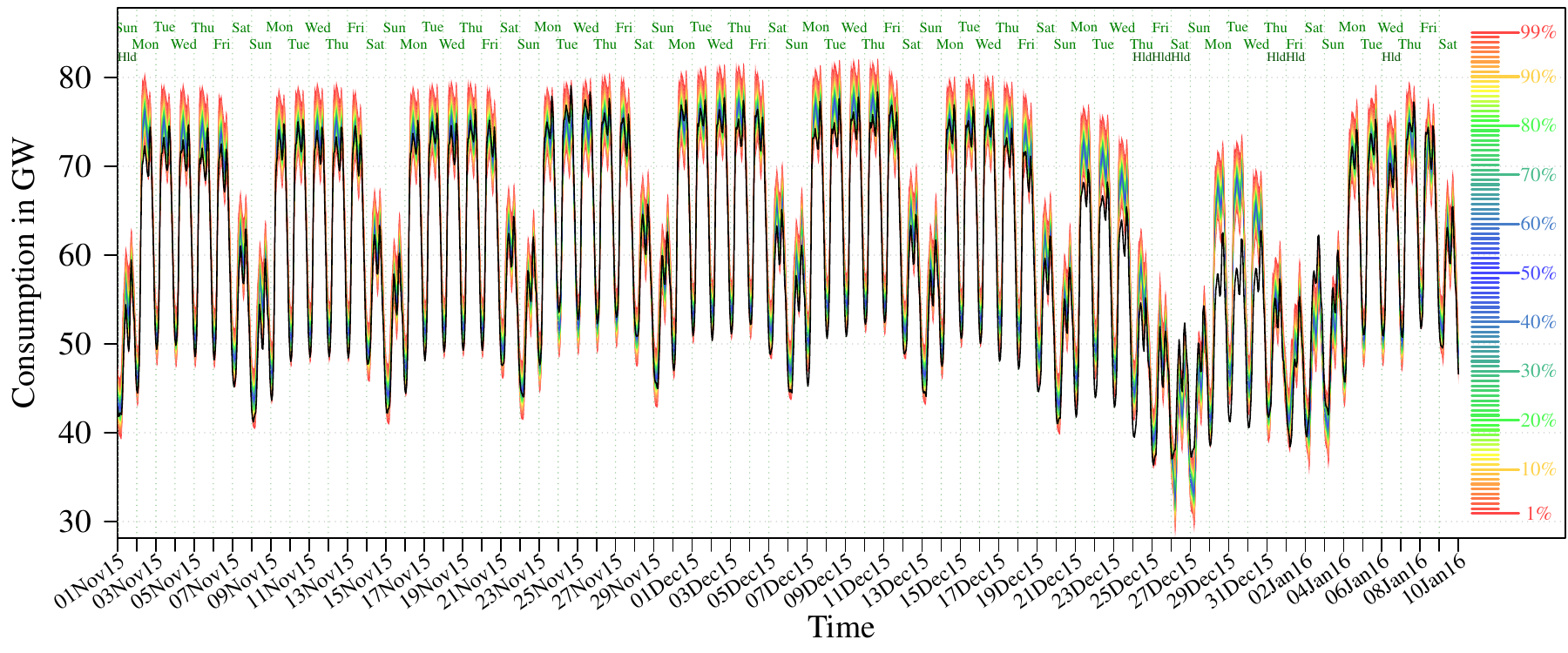}
\caption{Consumption forecast}
\label{fig_consumption_prediction_bands}
\end{subfigure}
\begin{subfigure}[b]{.99\textwidth}
 \includegraphics[width=1\textwidth]{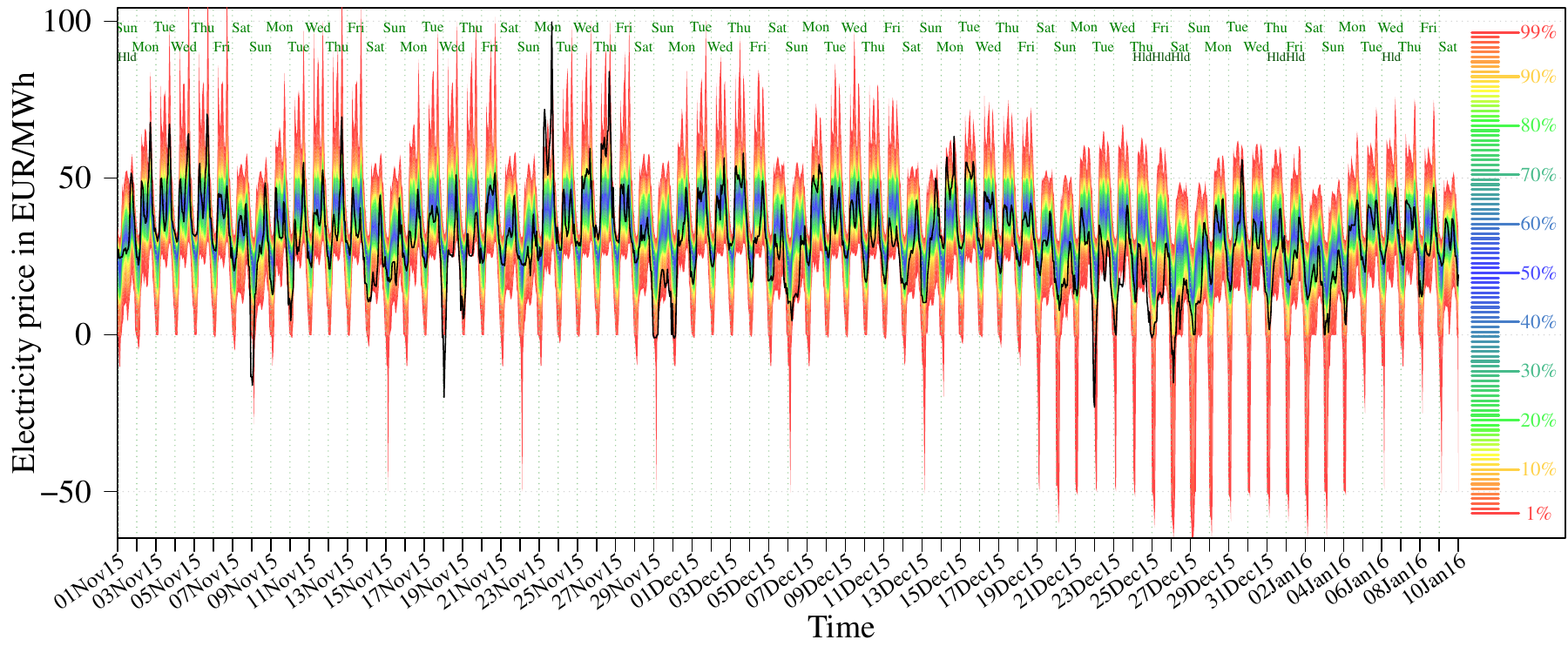}
\caption{Price forecast}
\label{fig_price_prediction_bands}
\end{subfigure}
\caption{Forecasts for the electricity consumption and the electricity price for the 99 percentiles (colored) with observed values (black) from 1 November 2015 to 10 January 2016.
The corresponding weekdays and public holidays are highlighted.}
\label{fig_prediction_bands}
 \end{figure}

 In Figure \ref{fig_prediction_bands} we show the probabilistic forecasting results for the electricity consumption and the electricity price.
The graphs show the 99 percentile estimates, e.g. the quantiles for 1\%, 2\%, $\ldots$, 99\%, for both processes from the 1 November 2015 to the 10 January 2016.
As the forecasts starts at the 20 April 2015 {this time frame corresponds to the seventh to ninth month of the forecasting horizon.}
Note that the forecasted winter period has usually a relatively high electricity consumption, except for the holiday period around Christmas and New Years Day.
In both graphs, \ref{fig_consumption_prediction_bands} and \ref{fig_price_prediction_bands}, the quantile estimates are displayed using colored prediction bands.
The median estimate (50\% quantile) is given by the blue colored center. Rare events, e.g. the 1\% and 99\% quantile, are reddish colored. The observed consumption and price values 
are added to the graphs using a black line. The day of the week as well as the considered (public) holidays are shown at the top of each picture.

For the consumption we see that the overall behavior is well captured. The estimated daily and weekly seasonalities match clearly the observed pattern.
However, at the working days around Christmas the consumption is overestimated. Still, on the holidays and weekends during this period it is relatively well captured, even though the available past information is limited.
Note that even the effect of the regional public holiday Epiphany on 6 January is appropriately estimated.

For the electricity prices in \ref{fig_price_prediction_bands} we observe similar daily and weekly seasonal patterns as for the consumption in \ref{fig_consumption_prediction_bands}. The electricity price tends to be smaller during night than during the day, and smaller on the weekend than during the standard weekday.
In general it seems that the overall price behavior is well captured, as most of the realized prices fall into the colored areas. 
We see that working days in November and beginning of December had a reasonable chance for a relatively high electricity price greater than 70 EUR/MWh. Indeed,
on Monday, the 23 November, and Thursday, the 26 November, prices larger than 70 EUR/MWh occurred.
Furthermore, we can see that the model for the electricity prices estimates a small chance for negative prices. The highest likelihood is during the Christmas and New Years Days holiday period.
Similarly, we observed negative prices for the 22 December and the 26 December. In general, the price model seems to struggle for the winter holiday period not as much as the 
consumption model as illustrated in \ref{fig_consumption_prediction_bands}. Still, the electricity prices tend to be overestimated within this period. However,
by eyeballing Figure \ref{fig_price_prediction_bands}, the long term electricity price forecasts seems to capture the relevant behavior well.

 As motivated in the introduction, renewable energy producers that get subsidies according to the EEG 2014 have a natural interest in forecasts of negative prices. 
As the subsidies get cut  if the electricity price is 6 times in a row negative they are interested in the probability of those events.
Therefore we call an event where the electricity price is negative for $c$ hours in a row, a $c$h-price$\leq$0 event. 
For instance, at time $t$ a 6h-price$\leq$0 event occurs if the electricity price at time $t$ and the five prices before $t$ are negative. 
As mentioned above we can easily estimate the probabilities of these events by evaluating relative frequencies of the $N$ events.
In Figure \ref{fig_prob_of_events} the probabilities for the 1h-price$\leq$0 and 6h-price$\leq$0 events are presented for the time period of the 1 November 2015 to the 10 January 2016.
For comparison purpose this is the same time range as used in Figure \ref{fig_prediction_bands}.
In this figure the 24 different hours when such an event occurred are highlighted by different colors. 

  \begin{figure}[hb!]
 \begin{subfigure}[b]{.99\textwidth}
 \includegraphics[width=1\textwidth]{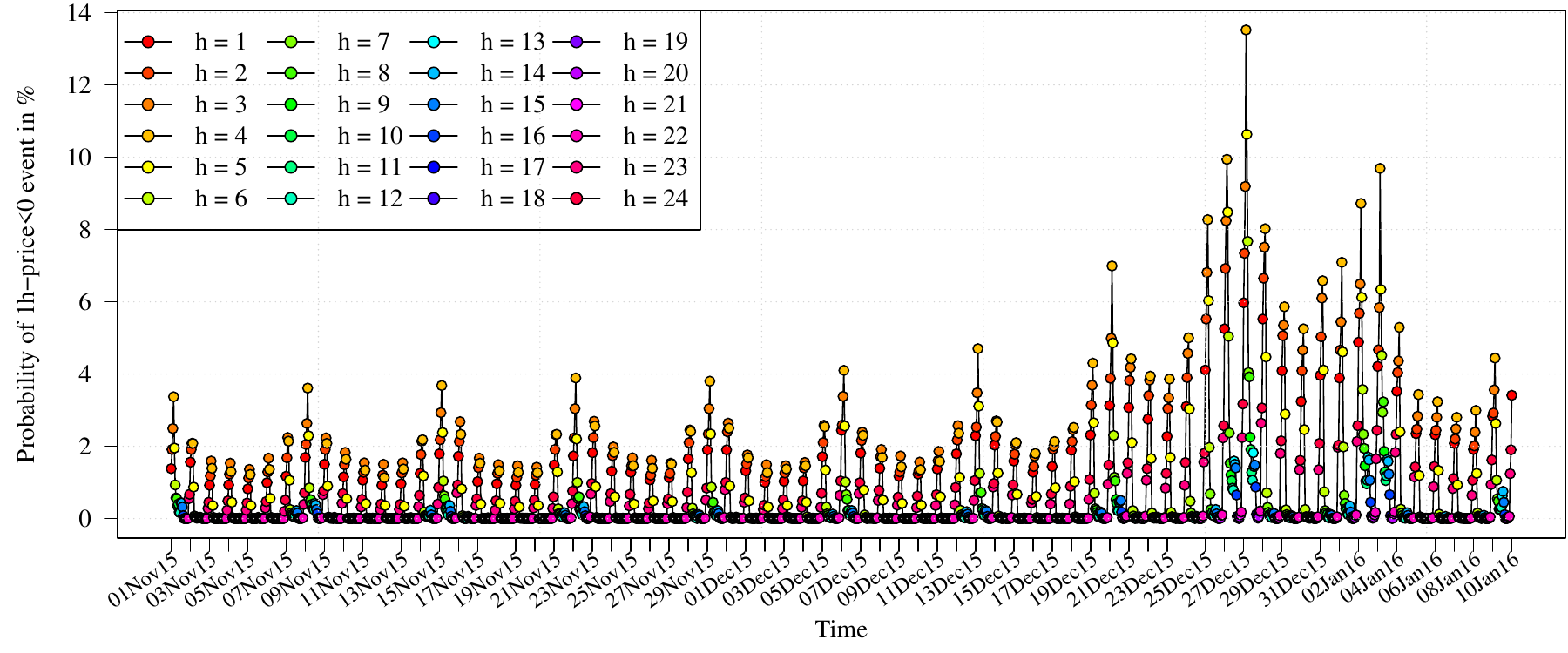}
\caption{$c=1$}
\label{fig_event_h=1}
\end{subfigure}
\begin{subfigure}[b]{.99\textwidth}
 \includegraphics[width=1\textwidth]{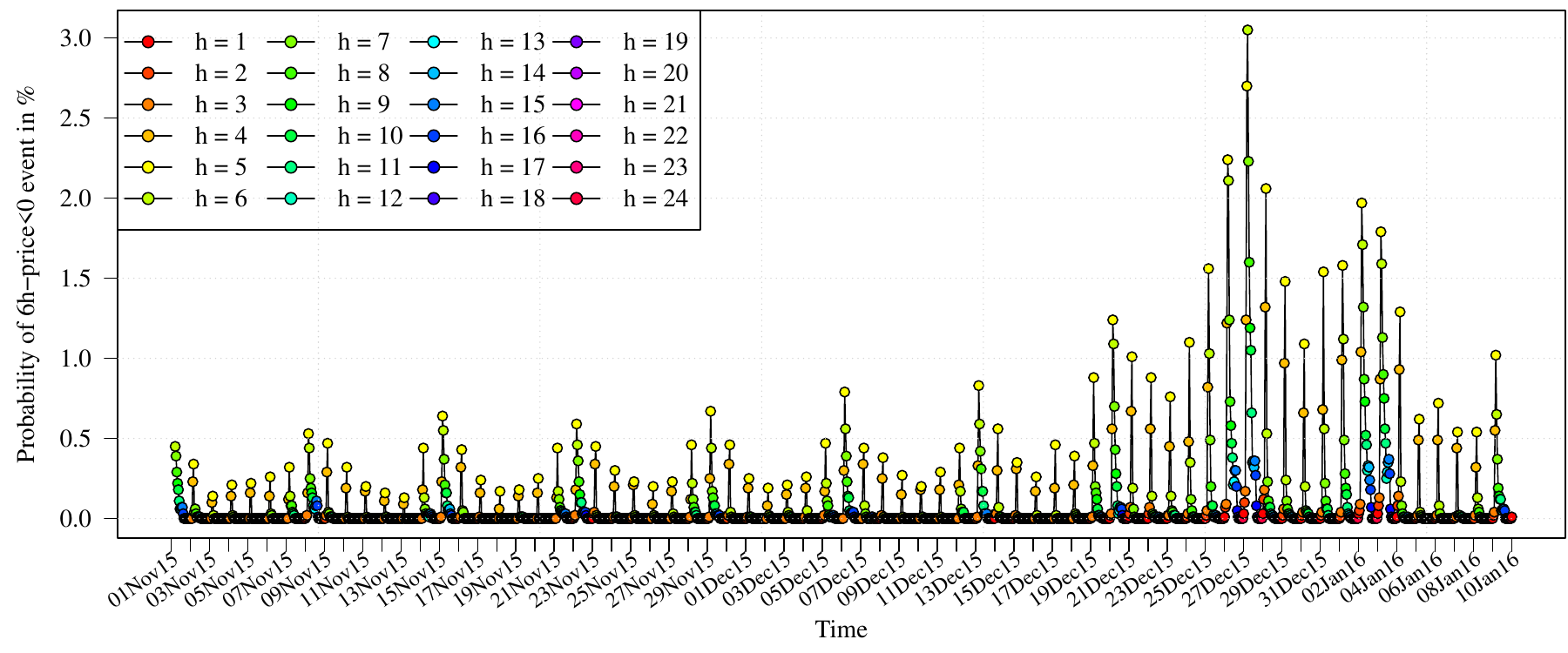}
\caption{$c=6$}
\label{fig_event_h=6}
\end{subfigure}
\caption{Forecasted probabilities of $c$h-price$\leq$0 events.}
\label{fig_prob_of_events}
 \end{figure}
 
We observe that in general the probabilities for these negative price events are small. Obviously, the probability for 1h-price$\leq$0 events is larger than for 6h-price$\leq$0 events.
For 6h-price$\leq$0 events the probability is estimated always below 3\% and usually even below 1\%. 
We observe that the highest probabilities for such events usually occur during the early morning hours $h=4$, $h=5$ and $h=6$. These probabilities are even higher on Monday's early morning. In contrast, the probability for one negative price, e.g. an 1h-price$\leq$0 event, is usually the highest during the night hours $h=24$, $h=1$, $\ldots$, $h=5$.
{This finding can be explained by the fact that for 6h-price$\leq$0 events the first five prices need to be negative and as we receive that prices start being negative at hours around $h=24$ the first consecutive six hours of negative prices can be found at around $h=5$, which matches our finding for the 6h-price$\leq$0 events.
Furthermore, we see in Figure \ref{fig_event_h=1} and \ref{fig_event_h=6} that the probabilities for 1h-price$\leq$0 and 6h-price$\leq$0 events are larger on weekend nights 
than during the working week. Another very distinct observable feature is that during the winter holiday period the probability is clearly higher than during weeks for both events. The 27 December 2016, which
is a Sunday followed by the Christmas holidays, seems to exhibit the highest probability for these negative price events.}
By comparing the probabilities for negative price events with the real prices in Figure \ref{fig_price_prediction_bands} we can conclude that during the days for which the probabilities seem to increase are also the days which had the most amount of negative prices in reality.
 
Note, that at the time when the study is conducted we only had real data up to the 30 November 2016 available, as the remaining time-frame of the 3-year ahead forecasts is still in the future. Thus, we do any following evaluation based on the first 1.5 years {(about 13000 out-of-sample observations)} of our forecasts. 

{For the purpose of comparison we add the forecasting results of a simple but very competitive benchmark for short- and mid-term forecasting, as used in \cite{ziel2015efficient} and \cite{ziel2016day}.
This forecasting model is the autoregressive time series model (\textbf{AR-HoW}) with weekly mean, formally defined as
\begin{equation}
 Y_t = \mu_t + \sum_{k=1}^p \phi_k (Y_{t-k} - \mu_t) + \eps_t.
\end{equation}
where $\mu_t$ is the weekly mean, which we estimate by the sample mean for each hour of the week. 
The autoregressive parameters $\phi_k$ are estimated by solving the Yule-Walker equations.
The order $p$ is selected between $1,\ldots,p_{\max}=2520$ (= potential memory of 15 weeks) such that the Akaike information criterion (AIC) is minimized. 
We simulate the residuals non-parametrically using bootstrap.
}

In Table \ref{tab_prob_events} we see the average {forecasted probabilities, measured as relative frequencies, of the \textbf{X-Model} and the \textbf{AR-HoW} of 
$c$h-price$\leq$0 events} 
for the out-of-sample range from 20 April 2015 to 30 November 2016 as well as the observed relative frequency of negative price events.
We see that overall the forecasted probability for a negative price is about $0.77\%$ {for the \textbf{X-Model} whereas}  the observed prices exhibit a relative frequency of $0.88\%$ which is {with a relative error of about 10\%} remarkably close for such a long forecasting horizon. {In contrast the \textbf{AR-HoW} has a forecasted probability of $1.61\%$
for the $1$h-price$\leq0$ event. This is about $80\%$ higher than the observed relative frequency.}
Obviously, the forecasted probabilities and their observed counterpart decrease in $c$. For the important $6$h-price$\leq$0 event {the \textbf{X-Model}} forecasted an average probability of about $0.09\%$ but observed this event with a probability of $0.16\%$, which is still quite good for a complex event and long forecasting horizons.
\begin{table}[ht]
\centering
{
\begin{tabular}{rrrrrrr}
  \hline
$c$ & 1 & 2 & 3 & 4 & 5 & 6 \\ 
  \hline
   \textbf{X-Model} & 0.768 & 0.523 & 0.356 & 0.232 & 0.143 & 0.087 \\ 
   \textbf{AR-HoW} & 1.607 & 1.111 & 0.774 & 0.545 & 0.386 & 0.275 \\ \hline
\textbf{observed} & 0.877 & 0.651 & 0.488 & 0.354 & 0.241 & 0.156 \\ 
   \hline
\end{tabular}
}
\caption{Observed and forecasted probabilities of $c$h-price$\leq$0 events in $\%$ for the out-of-sample range from 20 April 2015 till 30 November 2016.}
\label{tab_prob_events}
\end{table}

 We have seen in Table \ref{tab_prob_events} that the observed frequency of negative prices is close to the forecasted probabilities of the considered model in the out-of-sample range.
 However, by this table it cannot be concluded that such an event happened when we actually forecasted a high probability. 
 {Therefore we created Figure \ref{fig_histprob}, which provides information on the relative frequencies of negative prices for the $1$h-price$\leq$0 event plotted against its forecasted probabilities, categorized in groups. For instance, there is a group of hours where the forecasted probability for negative prices ranged from about 9.1\% to 10.9\%. 
 So we expect that in total about 10\% of the hours that fall into that category had a negative price and 90\% did not have a negative price. If our probabilities were correct, the red bar, which represents the relative frequency of prices within that category which were not negative, would be exactly 0.9 and the stacked green bar would start from 0.9 and end at 1.0, representing 10\% of the whole distance. For the 9.1\% to 10.9\% group this is almost the case, which means that the probabilities seem to be properly forecasted. In Figure \ref{fig_histprob} we also added a black line, which represents the theoretical relationship of relative frequencies with the forecasted probabilities if all estimations were perfect. In this case, the center of the top of the red bars would be the points through which the line runs. The blue dashed line however is the line which provides the true fit given our forecasted probabilities, estimated via {weighted OLS regression
 where the weights are chosen as number of observations within each group.} 
 The high $R^2$ value of $91.4\%$ shows a high forecasting prediction power of the model for these events. Overall it can be stated, that the higher the forecasted probability for a specific price, the more likely it was for this price to exhibit a negative value, exhibiting an almost perfect relationship.}

 \begin{figure}[htb!]
 \begin{subfigure}[b]{.99\textwidth}
 \includegraphics[width=1\textwidth]{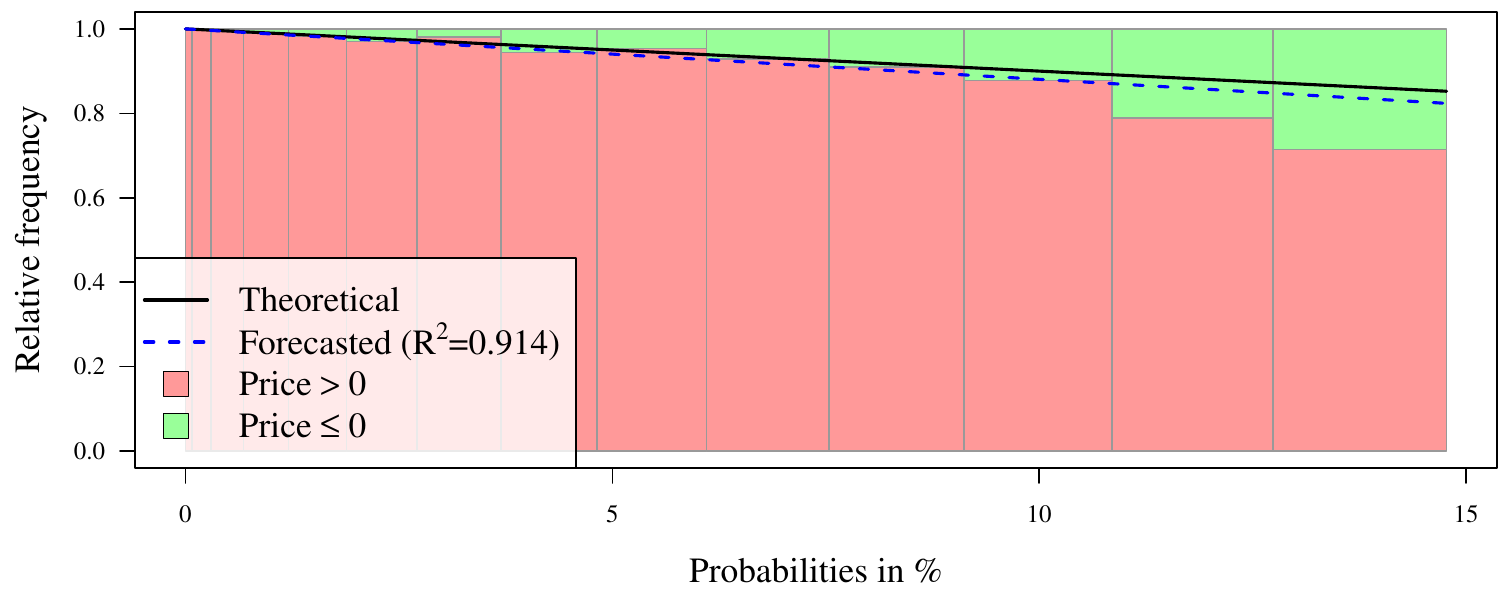}
\caption{Results for \textbf{X-Model}}
\label{fig_histprob_Xmodel}
\end{subfigure}
 \begin{subfigure}[b]{.99\textwidth}
 \includegraphics[width=1\textwidth]{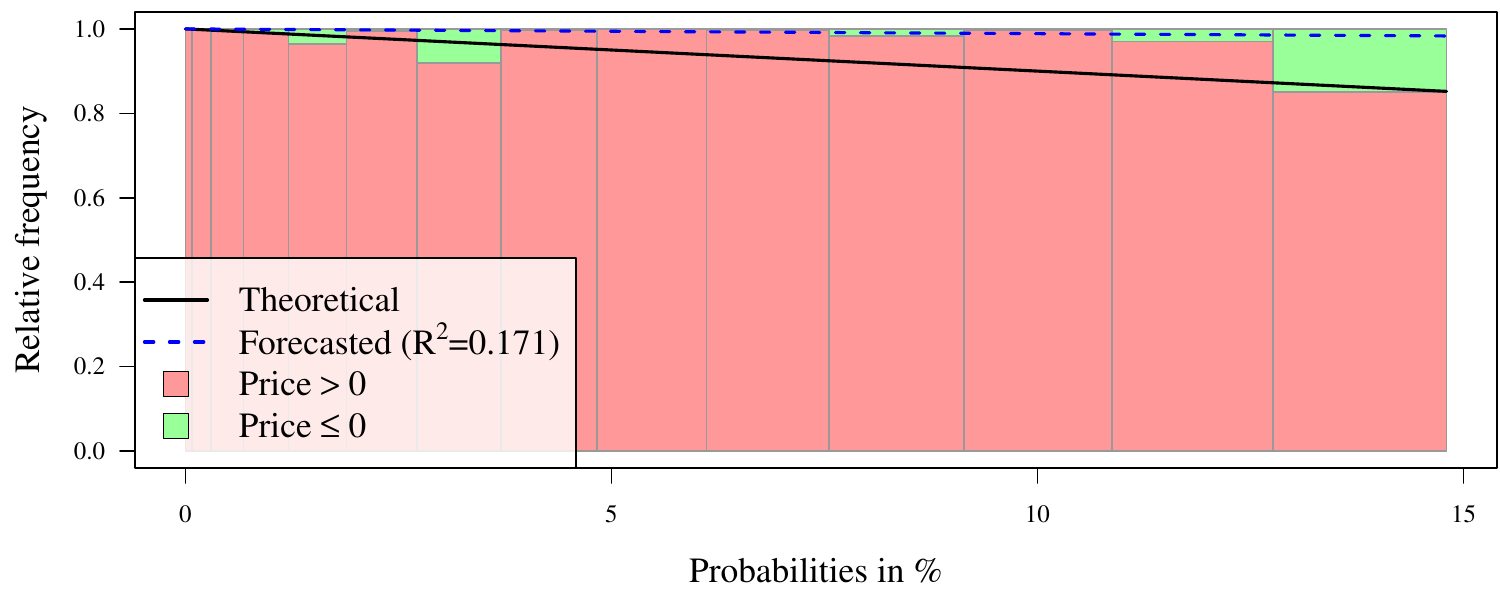}
\caption{Results for \textbf{AR-HoW}}
\label{fig_histprob_ARHoW}
\end{subfigure}

\caption{Relative frequencies of observed negative prices ($1$h-price$\leq$0 event) for different forecasted probabilites {of the X-Model and the AR-HoW}.}
\label{fig_histprob}
 \end{figure}

{As the X-model is tailor-made for probabilistic forecasting, it is {suitable} to perform a statistical based evaluation of the forecasting accuracy for the whole prediction interval.}
 Therefore, we evaluate the coverage probability of the quantile estimates. In detail we consider the 99 percentile estimates as used in Figure \ref{fig_prediction_bands}.
\begin{figure}[hb!]
 \begin{subfigure}[b]{.99\textwidth}
 \includegraphics[width=1\textwidth]{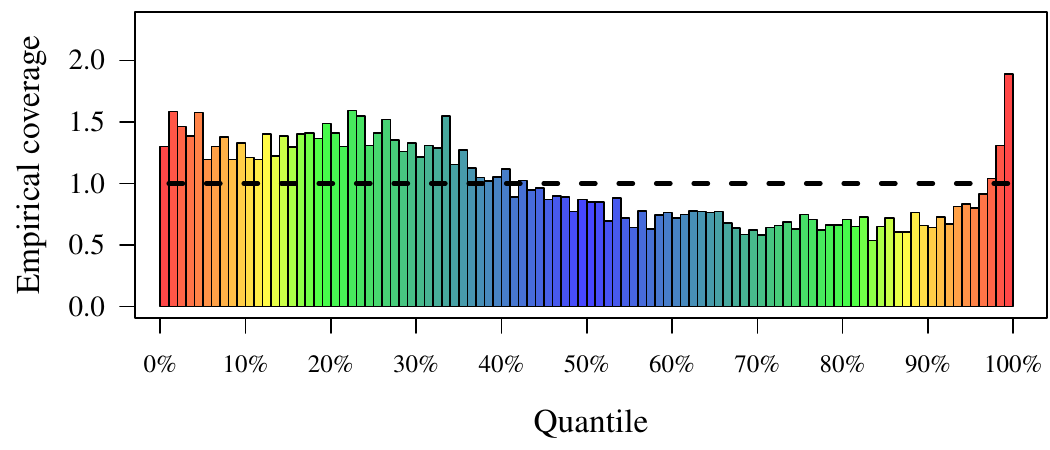}
\caption{Results for \textbf{X-Model}}
\label{fig_pit_Xmodel}
\end{subfigure}
 \begin{subfigure}[b]{.99\textwidth}
 \includegraphics[width=1\textwidth]{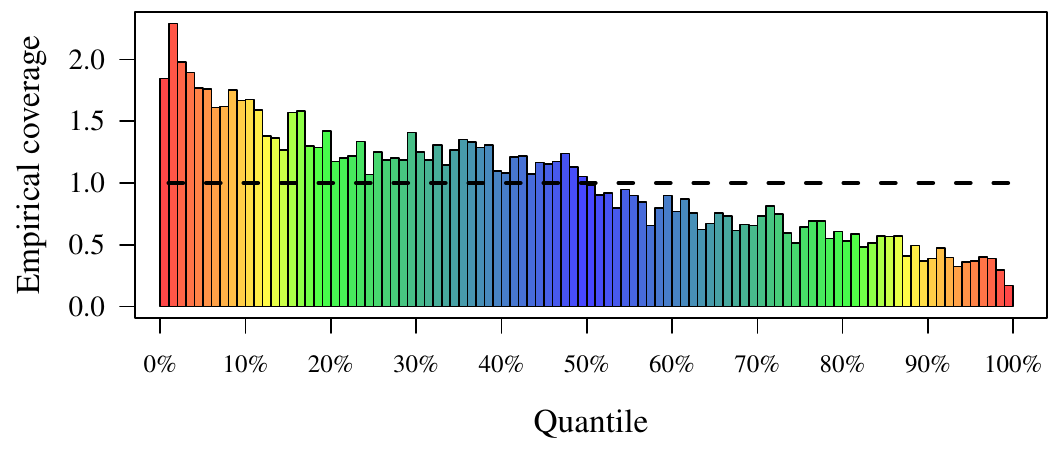}
\caption{Results for \textbf{AR-HoW}}
\label{fig_pit_ARHoW}
\end{subfigure}
\caption{Pit plot with uniform distribution as benchmark as dashed line {for the X-Model and the AR-HoW.}}
\label{fig_pit}
 \end{figure}
 The estimated coverage probabilities of the full available time range is given in Figure \ref{fig_pit}. The colors of the bars in the histogram \ref{fig_pit} match the prediction band colors of Figure \ref{fig_prediction_bands} for easier comprehension. {Each bar represents the relation of how many real values were in that specific estimated quantile to the amount of theoretical values that should be in that quantile. This means that if we have an empirical coverage of more than 1 for a quantile we can conclude that in reality more prices fell into this estimated quantile than expected.} In the optimal case the colored distribution in the histogram should not be distinguishable from the uniform distribution
 which is highlighted by the dashed line. 
 Clearly this is not the case here, which indicates that some systematical errors may still be present in the modeling approach.
The lower quantiles till 35\% seem to be overrepresented by around 30\%, which means that our model underestimated the probability for low prices. The quantiles around 50\% to 95\% 
seem to be underrepresented which means that our model estimated the likelihood of higher prices too high.
 Similarly as in the short term model in \cite{ziel2015efficient} the extreme upper quantiles (98\% and 99\%) are overrepresented as well. {However, as we are comparing the forecasting results of around 1.5 years ahead with hourly resolution, we still consider these results as promising.}

{Nevertheless, our modeling approach still leaves room for improvements.} For instance, adjustment methods as discussed in e.g. \cite{bello2016parametric} for electricity price forecasting may help to reduce the bias even further.

 From the modeling perspective there are many areas where single model components can be improved and increase the overall accuracy, 
 this holds for the statistical modeling perspective but also for the fundamental model perspective. 
 Also the capacity assumptions can be improved, additional plans for specific shut-downs of other power plants could be incorporated into the model. Moreover, 
 other economic variables, like fuel prices, especially coal and natural gas as well as $\text{CO}_2$-costs or GDP-growth can be integrated into the model to capture associated dependencies and uncertainty.
 Moreover the X-model itself that is used to describe the bidding behavior relationships can be improved. Similarly we could incorporate price information from the future market into the day-ahead market expectations for the future.
 Finally the market coupling and cross-border aspects as well as political risks were excluded from the model. An incorporation of these effects could likely improve the overall performance. 
 {A possible way of implementing these effects may be done via scenario analysis, where different target values of governments policy plans are considered.}

\section{Conclusion} \label{conclusion}

In this research study we proposed an innovative modeling methodology for mid- and long-term forecasting of electricity prices.
The approach includes a complete modeling of the electricity market structure, including modeling components like electricity consumption and power generation of renewable energy like wind and solar. Furthermore, it takes the detailed day-ahead market bidding structure into account by modeling the market bids for the supply and demand side. 

{By an extensive literature review about different forecasting horizons and probabilistic electricity price forecasting we detected that there was no single article published which considered long-term forecasting as well as probabilistic forecasting within its model approach. However our} results show that the application of path based electricity price forecasting is possible even in the long run. 
By simulating many possible physical market situations with the resulting day-ahead expectations and the corresponding bids on the day-ahead market we are able to get a realistic map of the future electricity prices. This includes not only the mean and variance behavior, but also the path dependent interactions. We show that the estimation of probabilities of complex events even far in the future is feasible. {Regarding this, we provide information on probabilities for the occurrence of at least one negative price and the practical relevant event of six consecutive negative prices. 
{Due to the comprehensive modeling of the physical market situation, it is also possible to implement market instabilities like outages of specific power plants or extreme weather events. The X-model we used can easily map their impacts into the market relevant sell and purchase curves. So if an agent who uses this model is certain about a specific event to happen, it could be easily included into the forecasting results. However, due to the simulation strategy we pursued, these instabilities are automatically and randomly included as the model has learned from such events in the past.}

As we are able to forecast them up to three years in advance, this may help practitioners for their investment decisions, e.g. concerning the construction of new wind power plants. {Another more trading based application could be the detection of possible speculative opportunities. As the EEX also offers a wide range of long term derivatives, e.g. futures, our model could provide a decision support for the important decision of buying a future or wait until the settlement period has arrived and buy the spot product at that time instead.} By forecasting the whole probabilistic density of prices we can support the risk management of electricity companies in their decision making even in the long-run. 

A backtesting study for about 1.5 years showed that our model is able to capture the full behavior of the electricity price quite well, given the long time horizon. We further demonstrate, that econometric models are eligible to long-term forecasting horizons, if the whole market situation is considered in the model structure. Therefore, we reduce the gap between econometric modeling approaches and fundamental market approaches even further.}

{
\section{Acknowledgement}

We thank the European Power Exchange (EPEX SPOT SE), European Network of Transmission System Operators (ENTSO-E), German Meteorological Office (DWD)
and the German Federal Network Agency (BNetzA) for providing the datasets we have used in this article.}

\bibliographystyle{apalike}
\bibliography{x-model_long}

\end{document}